\documentclass[usenames,twocolumn,dvipsnames]{aastex63} 
\usepackage[T1]{fontenc}
\usepackage{graphicx}
\usepackage{dcolumn}
\usepackage{bm}

\usepackage{amsmath}	
\usepackage{amssymb}	
\usepackage{txfonts}
\usepackage[bb=libus]{mathalpha}
\usepackage{bbold}
\usepackage{relsize}
\usepackage{mathtools}
\usepackage{bigints}
\usepackage{sidecap}
\usepackage[caption=false]{subfig}
\usepackage{hyperref}

\usepackage{floatrow}
\usepackage{soul}
\usepackage{comment}
\usepackage[normalem]{ulem}

\usepackage{macros_ub}
\graphicspath{{./}{figures/}}

\received{XXX}
\revised{YYY}
\accepted{ZZZ}

\begin{document}
\vspace{1mm}


\shorttitle{Quasilinear relaxation of collisionless plasmas}
\shortauthors{}

\title{Universal non-thermal power-law distribution functions from the self-consistent evolution of collisionless electrostatic plasmas}

\correspondingauthor{Uddipan Banik}
\email{uddipan.banik@princeton.edu,\;uddipanbanik@ias.edu}

\author[0000-0002-9059-381X]{Uddipan Banik}
\affiliation{Department of Astrophysical Sciences, Princeton University, 112 Nassau Street, Princeton, NJ 08540, USA}\affiliation{Institute for Advanced Study, Einstein Drive, Princeton, NJ 08540, USA}\affiliation{Perimeter Institute for Theoretical Physics, 31 Caroline Street N., Waterloo, Ontario, N2L 2Y5, Canada}

\author[0000-0001-6411-0178]{Amitava Bhattacharjee}
\affiliation{Department of Astrophysical Sciences, Princeton University, 112 Nassau Street, Princeton, NJ 08540, USA}

\author[0000-0002-3415-2067]{Wrick Sengupta}
\affiliation{Department of Astrophysical Sciences, Princeton University, 112 Nassau Street, Princeton, NJ 08540, USA}


\begin{abstract}

Collisionless systems often exhibit non-thermal power law tails in their distribution functions. Interestingly, collisionless plasmas in various physical scenarios (e.g., the ion population of the solar wind), feature a $v^{-5}$ tail in the velocity ($v$) distribution, whose origin has been a long-standing puzzle. We show this power law tail to be a natural outcome of the collisionless relaxation of driven electrostatic plasmas. Using a quasilinear analysis of the perturbed Vlasov-Poisson equations, we show that the coarse-grained mean distribution function (DF), $f_0$, follows a quasilinear diffusion equation with a diffusion coefficient $D(v)$ that depends on $v$ through the plasma dielectric constant. If the plasma is isotropically forced on scales larger than the Debye length with a white noise-like electric field, $D(v)\sim v^4$ for $\sigma<v<\omega_{\mathrm{P}}/k$, with $\sigma$ the thermal velocity, $\omega_{\mathrm{P}}$ the plasma frequency and $k$ the characteristic wavenumber of the perturbation; the corresponding quasi-steady state $f_0$ develops a $v^{-(d+2)}$ tail in $d$ dimensions ($v^{-5}$ tail in 3D), while the energy ($E$) distribution develops an $E^{-2}$ tail independent of dimensionality. Any redness of the noise only alters the scaling in the high $v$ end. Non-resonant particles moving slower than the phase-velocity of the plasma waves ($\omega_{\mathrm{P}}/k$) experience a Debye-screened electric field, and significantly less (power-law suppressed) acceleration than the near-resonant particles. Thus, a Maxwellian DF develops a power-law tail, while its core ($v<\sigma$) eventually also heats up but over a much longer timescale. We definitively show that self-consistency (ignored in test-particle treatments) is crucial for the emergence of the universal $v^{-5}$ tail.

\end{abstract}


\section{Introduction}

The pursuit of universal velocity distribution functions for N-body systems has been a holy grail of kinetic theory since the remarkable results obtained by Ludwig Boltzmann in the late nineteenth century. It is well known that short-range interactions or collisions drive the velocity distribution function (DF) of a system (e.g., a neutral gas or a plasma) towards a Maxwellian. This can be understood in a number of ways: (1) the Maxwellian DF annihilates the collision operator in the Boltzmann (or the Fokker-Planck) equation that describes the collisional relaxation of a system, and is thus a steady-state solution, and (2) it is the DF that maximizes the Boltzmann-Shannon entropy, according to the Boltzmann H-theorem. The collision-driven decorrelation of the momenta of particles, also known as molecular chaos, is at the heart of the Boltzmann H-theorem, and ultimately leads to the irreversible upward march of the Boltzmann-Shannon entropy towards its maximum and the consequent establishment of the Maxwellian DF in thermal equilibrium. The ubiquity of the Maxwellian DF in collisional systems is a testament to its universal nature. 

Almost equally ubiquitous is the presence of power-law tails in non-thermal DFs in collisionless systems that are governed by long-range forces. On timescales over which they are observed, such systems do not equilibrate or relax via collisions, i.e., do not attain the maximum Boltzmann-Shannon entropy state. Yet, non-thermal DFs with power-law tails tend to be long-lived and represent a quasi-steady state. One specific power-law, the $v^{-5}$ tail in the three-dimensional (3D) velocity ($v$) distribution or the $E^{-2}$ tail in the energy ($E$) distribution, conspicuously appears in collisionless plasmas. For example, the suprathermal ion distribution in the solar wind is known to harbor a preponderance of $v^{-5}$ tails \citep[][]{Gloeckler.03,Simunac.Armstrong.04,Fisk.Gloeckler.12,Fisk.Gloeckler.14,Maksimovic.21}. In the inner heliosphere, the $v^{-5}$ tail in the DF of ions from H through Fe has been measured at non-relativistic energies (below $100$ keV - $1$ MeV/nucleon) by the Solar Wind Ion Composition Spectrometer (SWICS) instruments on the Ulysses and ACE spacecrafts \citep[][]{Gloeckler.etal.92,Fisk.Gloeckler.12}, and in the heliosheath, by the Voyager Low-Energy Charged Particle Experiment and Cosmic Ray Subsystem instruments over their entire energy range \citep[][]{Krimigis.etal.77,Stone.etal.77}. The $v^{-5}$ tail in the ion distribution was also observed by Voyagers 1 and 2 soon after they crossed the termination shock \citep[][]{Decker.etal.06a,Decker.etal.06b,Gloeckler.etal.08}. The origin of the preponderance and persistence of this power-law tail has been a subject of long-standing interest and controversy. Similar non-thermal tails have also been observed in the kinetic ion distribution of high energy density thermonuclear plasmas generated by laser driven inertial confinement fusion implosions \citep[][]{Mannion.etal.23}. Although the approach pursued in this paper is not directly applicable to magnetized plasmas, the fundamental principles are applicable and will be the subject of future work.  

A collisionless electrostatic plasma is described by the collisionless Boltzmann or the Vlasov equation and the Poisson equation (in the non-relativistic regime). It is well known that the Vlasov equation admits a denumerably infinite set of Casimir invariants (of which the Boltzmann H-function is but one); any positive definite function of the conserved quantities is a steady-state solution to the Vlasov equation. Why then does a collisionless system tend to relax to a particular quasi-steady state? This is because, while the fine-grained DF obeys the Vlasov equation, in numerical experiments or satellite observations one typically measures the coarse-grained DF, which is some averaged version of the fine-grained DF. This coarse-grained DF does not
follow the Vlasov equation but a modified kinetic equation, with an effective collision operator (in a mean-field theory) that represents the effect of instabilities and/or turbulence. The effective collision operator obtained by the coarse-graining of collisionless plasmas is, in general, quite different than the Boltzmann collision operator. Hence, the maximum entropy state (if such a thing even exists in this case) can exhibit significant non-Maxwellian
features.

It is argued by some that the effective collision operator is of the Balescu-Lenard form \citep[][]{Ewart.etal.22}. Interestingly, \cite{Ewart.etal.23} have shown that maximizing a modified form of the Boltzmann-Shannon entropy, where they treat the DF as a random variable and replace the DF in the entropy expression by the probability that the DF takes a certain value (inspired by \citet[][]{LyndenBell.67}), yields the $E^{-2}$ distribution, hinting at the universality of this power-law. Others have argued for the ubiquity of kappa distribution functions as a replacement for the Maxwellian from a novel statistical mechanics of collisionless systems (see \cite{Livadiotis.McComas.13} and other references therein). 

The primary goal of this paper is to develop a fully self-consistent evolution equation for the mean, coarse-grained DF, $f_0$, of an externally driven collisionless electrostatic plasma. Our main tool is quasilinear theory (QLT) (second order perturbation theory; see \citet[][]{Diamond_Itoh_Itoh_2010} for a modern treatment). Of course, QLT comes with its own set of assumptions, some of which are not universally applicable. The fundamental assumption is that the problem of collisionless relaxation admits a separation of timescales, i.e., the mean, coarse-grained DF evolves over a timescale much longer than the plasma oscillation period or the typical timescale associated with linear fluctuations. In the violent relaxation \citep[][]{LyndenBell.67} of collisionless plasmas, standard QLT does not describe the evolution of coherent structures such as Bernstein-Green-Kruskal \citep[][]{Bernstein.etal.57} (BGK) modes/holes, where the nonlinear trapping/libration time of charged particles is of order only a few plasma oscillation periods. In other words, QLT does not describe the evolution of the fluctuations \citep[see][for a (non self-consistent) study]{Nastac.etal.23}, which can be subject to higher order non-linear effects. However, QLT does describe with reasonable accuracy the evolution of the mean coarse-grained DF of the bulk plasma, even in unstable situations such as two-stream instabilities \citep[][]{Ewart.etal.24}. In fact, the $E^{-2}$ power-law tail in $f_0$ that arises from our quasilinear formalism is the same as what 1D PIC simulations of two-stream instabilities by \citet[][]{Ewart.etal.24} predict. The precise reason why QLT seems to work (to a reasonable degree) even in strongly nonlinear problems is unclear and remains an open question.

QLT yields an evolution equation of the Fokker-Planck type for the mean DF, with a diffusion coefficient that depends on the plasma dielectric constant. We demonstrate that if the plasma is driven by a white noise-like stochastic electric field isotropically on scales larger than the Debye length ($k\lambda_\rmD \ll 1$), the diffusion coefficient scales universally as $\sim v^4$ for velocities between $\sigma$, the velocity dispersion of the DF, and $\omega_\rmP/k$, the phase-velocity of the plasma waves, with $k$ the characteristic wavenumber of the external field and $\omega_\rmP$ the plasma frequency (or ion sound frequency). This ultimately establishes a $v^{-\left(d+2\right)}$ power-law tail in the quasi-steady state DF, with $d$ the number of dimensions, i.e., a $v^{-5}$ tail in 3D. This corresponds to an $E^{-2}$ tail in the energy distribution, irrespective of the dimensionality of space. We demonstrate that the presence of temporal correlations in the noise (red noise) partially breaks this ``universality" and modifies the power-law exponent for velocities larger than $1/k\tau_\rmc$, where $\tau_\rmc$ is the noise correlation time.

To explain the origin of the $v^{-5}$ tail in the ion population of the solar wind, Fisk \& Gloeckler, in a series of papers \citep[][]{Fisk.Gloeckler.06,Fisk.Gloeckler.07,Fisk.Gloeckler.08,Fisk.Gloechler.09,Fisk.etal.10,Fisk.Gloeckler.12,Fisk.Gloeckler.14,Gloeckler.Fisk.06}, perform a quasilinear treatment of the Parker transport equation for the evolution of the DF in the solar wind frame. They argue that particles from the core of a Maxwellian distribution can be accelerated to high energies, forming a non-thermal $v^{-5}$ tail, by adiabatic compressions and expansions driven by the solar wind, accompanied by diffusion (something they call the `pump mechanism'). However, \cite{Jokipii.Lee.10} argue that the treatment by Fisk \& Gloeckler does not conserve particle number, and that a proper treatment of their proposed mechanism of stochastic acceleration can only yield power-law tails shallower than $v^{-3}$. 

It should be noted that neither Fisk \& Gloecker nor Jokipii \& Lee are fully self-consistent models (this limitation is explicitly acknowledged in \cite{Jokipii.Lee.10}). The ions are treated as test particles, but the self-consistent coupling of the fields to the ion DF through Maxwell's equations, which would provide a back-reaction on the DF, are not included in their treatments. In contrast, within the range of validity of the electrostatic approximation which includes the self-consistent Poisson equation, we demonstrate that the DF exhibits a universal $v^{-5}$ tail. We thus conclude that the requirement of self-consistency imposes powerful constraints on the form of DFs, and when such constraints are included, one can obtain results on the universality of power-law tails. The reason for this conclusion in the present context is not hard to see. Charged particles in the plasma do not see the bare electric field, rather they experience the `dressed' field and are themselves dressed due to Debye shielding. This implies that the non-resonant particles in a DF, moving slower than the phase-velocity ($\omega_\rmP/k$) of the plasma waves, are more screened, acquire an effective charge much smaller than the bare charge and are therefore less accelerated than the near-resonant particles ($v\sim \omega_\rmP/k$), causing the quasilinear diffusion coefficient to develop a $v^4$ dependence and the quasi-steady state DF to develop a $v^{-5}$ tail for $\sigma < v < \omega_\rmP/k$. Besides the self-consistency requirement, the development of this power-law tail requires the following conditions: (1) isotropic electrostatic forcing on scales much larger than the Debye length (e.g., in Langmuir or ion sound turbulence), and (2) white noise-like (small correlation time) forcing. Red noise with correlation time $\tau_\rmc\gtrsim 1/\omega_\rmP$ modifies the power-law exponent for high velocity particles with $v>1/k\tau_\rmc$. Forcing on scales comparable to the Debye length and anisotropic forcing also modify the power-law. This may explain why, despite the preponderance of $v^{-5}$ tails in the solar wind data, there exist parts of the phase-space that show deviations from it \citep[][]{Jokipii.Lee.10}.

This paper is organized as follows. Section~\ref{sec:lin_resp_theory} introduces the perturbative (linear and quasilinear) response theory for the relaxation of driven collisionless plasmas governed by the Vlasov-Poisson equations, and derives the quasilinear Fokker-Planck/diffusion equation for the evolution of the mean coarse-grained DF. In Section~\ref{sec:QL_diff_coeff}, we discuss the properties, in particular the velocity scalings, of the quasilinear diffusion coefficient for different noise models. In section~\ref{sec:QL_DF}, we solve the quasilinear diffusion equation and obtain the universal velocity scaling of the mean coarse-grained DF in the quasi-steady state. We summarize our findings in section~\ref{sec:discussion_summary}.

\section{Perturbative response theory for collisionless plasmas}\label{sec:lin_resp_theory}

A plasma is characterized by the DF or phase space ($\bx,\bv$) density of particles, $f(\bx,\bv,t)$. In this paper, we shall restrict ourselves to studying the evolution of the DF of a single charged species with other species included as part of a charge-neutralizing background, but the treatment can be extended to include other species self-consistently (with significant additional complexity in the algebra). The general equations governing the evolution of a collisionless electrostatic plasma are the Vlasov equation and the self-consistent Poisson equation. The Vlasov equation,

\begin{align}
\frac{\partial f}{\partial t} + \bv\cdot{\bf \nabla} f + \frac{e}{m}{\bf \nabla}_\bv f \cdot \left(\bE^{(\rmP)}+\bE\right) = 0, 
\label{Vlasov_eq}
\end{align}
is a conservation equation for the DF, $f$, of the charged species under consideration. Here, $e$ is the electric charge (same as the electron charge for electrons and $-Z$ times the electron charge for ions with atomic number $Z$), $m$ is the mass of the charged species, $\bE$ is the self-generated electric field of the plasma that is sourced by the DF via the Poisson equation,

\begin{align}
\nabla\cdot\bE = \frac{e}{\epsilon_0} \int \rmd^d v\, \left(f-\sum_{\rms} f_\rms\right),
\label{Poisson_eq}
\end{align}
where $\epsilon_0$ is the permittivity of vacuum, $d$ is the number of dimensions, and $f_\rms(\bx,\bv,t)$ is the DF of each of the other charged species. We assume quasi-neutrality in equilibrium, i.e., the number density, $n_e$, of the species under consideration is equal to the total number density of the other charged species, $\sum_{\rms} n_\rms$. $\bE^{(\rmP)}$ is the perturbing electric field due to forcing ``external" to our system (to be made more precise below), that may be sourced by perturbations in $f_\rms$.

The dynamics of a collisionless electrostatic plasma is fully described by the above Vlasov-Poisson system of equations. These are difficult to solve in their full generality (due to the non-linearity of the Vlasov equation), and hence, one must resort to perturbation theory to obtain analytical solutions. If the strength of the perturber potential, $\Phi^{(\rmP)} = -\int \bE^{(\rmP)}\cdot \rmd \bx$, is smaller than $\sigma^2$, where $\sigma$ is the velocity dispersion of the unperturbed near-equilibrium system, then the perturbation in $f$ can be expanded as a power series in the perturbation parameter, $\epsilon \sim \left|\Phi^{(\rmP)}\right|/\sigma^2$, i.e., $f = f_0 + \epsilon f_1 + \epsilon^2 f_2 + ...\,$; $\bE$ can also be expanded accordingly.

As shown in Appendix~\ref{App:perturbation_theory}, one can perform a Fourier transform with respect to $\bx$ and Laplace transform with respect to $t$ of $f$, $\bE_1$ and $\bE^{(\rmP)}$, to derive the response of the system order by order. The Fourier-Laplace coefficients of the linear response can be summarized as follows:

\begin{align}
\Tilde{f}_{1\bk}(\bv,\omega) &= -\frac{ie}{m} \, \frac{\left(\Tilde{\bE}^{(\rmP)}_{\bk}(\omega) + \Tilde{\bE}_{1\bk}(\omega)\right)\cdot {\partial f_0}/{\partial \bv}}{\omega - \bk\cdot\bv},\nonumber\\
\Tilde{\bE}_{\bk}(\omega) &= \Tilde{\bE}^{(\rmP)}_{\bk}(\omega) + \Tilde{\bE}_{1\bk}(\omega) = \frac{\Tilde{\bE}^{(\rmP)}_{\bk}(\omega)}{\varepsilon_{\bk}(\omega)},\nonumber\\
\varepsilon_{\bk}(\omega) &= 1 + \frac{\omega^2_\rmP}{k^2} \int \rmd^d v\, \frac{\bk\cdot{\partial f_0}/{\partial \bv}}{\omega - \bk\cdot\bv},
\label{lin_resp_eq}
\end{align}
where $\omega_\rmP = \sqrt{n_e e^2/m\epsilon_0}$ is the plasma frequency (or the frequency of ion waves), $n_e$ being the number density of the charged species. The subscript $\bk$ stands for the Fourier transform in $\bx$ while the tilde represents the Laplace transform in $t$. The dielectric constant $\varepsilon_{\bk}$ represents the polarization of the medium and the consequent Debye shielding/screening of the electric field.

\subsection{Quasilinear theory}\label{sec:quasilinear}

The linear response $f_{1\bk}(t)$ (obtained by the inverse Laplace transform of $\Tilde{f}_{1\bk}(\omega)$ as shown in Appendix~\ref{App:perturbation_theory}) consists of a continuum response that evolves as $\exp{\left[-i\bk\cdot\bv t\right]}$ and a set of discrete Landau modes that evolve as $\exp{\left[-i\omega_{\bk n}t\right]}$, with the modal frequencies $\omega_{\bk n}$ (the subscript $n$ denotes the $n^{\rm th}$ discrete mode) following the Landau dispersion relation, $\varepsilon_{\bk}\left(\omega_{\bk n}\right)=0$. As shown by \cite{Landau.46}, these modes are oscillating but damped. On scales larger than the Debye length, $\lambda_\rmD = \sigma/\omega_{\rmP}$, Landau damping becomes inefficient and the plasma response consists of Langmuir waves oscillating at frequencies $\omega_{\bk}$ with $\omega^2_{\bk}\approx \omega^2_{\rmP}\left(1 + 3 k^2\lambda^2_\rmD\right)$ (the other branch consists of ion acoustic waves, which are excited if the ions are permitted to move.) Typically, the linear response evolves over a timescale of the order of the plasma oscillation period, $2\pi/\omega_\rmP$, which is much shorter than the evolution timescale of the mean DF.

The evolution of the mean DF averaged over the volume $V$ of the bulk plasma, $f_0 = {\left(2\pi\right)}^d f_{2\bk=0}/V$, can be studied by computing the second order response, $f_{2\bk}$, taking the $\bk \to 0$ limit and ensemble averaging the response over the random phases of the linear fluctuations (see Appendix~\ref{App:QLT} for details). This yields the following quasilinear equation \citep[see][for a comprehensive review]{Diamond_Itoh_Itoh_2010}:

\begin{align}
\frac{\partial f_0}{\partial t} &= -\frac{{\left(2\pi\right)}^d e}{m V} \int \rmd^d k \, \left<\bE_{\bk}^{\ast} \cdot \nabla_{\bv} f_{1\bk}\right>,
\label{quasilin_resp_eq}
\end{align}
where $\bE_{\bk} = \bE_{1\bk} + \bE_{\bk}^{(\rmP)}$. Here we have used the reality condition that $\bE_{-\bk}^{(\rmP)} = \bE_{\bk}^{(\rmP)\ast}$ and $\bE_{1,-\bk} = \bE_{1\bk}^{\ast}$. 

Now, we need to make assumptions about the temporal correlation of the external perturbing electric field, $E_{\bk i}^{(\rmP)}(t)$, where the subscript $i$ denotes the $i^{\rm th}$ component of $\bE_{\bk}^{(\rmP)}(t)$. We assume that $E_{\bk i}^{(\rmP)}(t)$ is a generic red noise:

\begin{align}
\left<E_{\bk i}^{(\rmP)\ast}(t) E_{\bk j}^{(\rmP)}(t')\right> &= \calE_{ij}\left(\bk\right)\,\calC_t\left(t-t'\right),
\label{white_noise_t}
\end{align}
where $\calC_t$ is the correlation function in time. This implies that $A_{\bk i}\left(\omega^{(\rmP)}\right)$, the Fourier transform of $E_{\bk i}^{(\rmP)}(t)$\footnote{Here, taking the Fourier transform in time is not very different from taking the Laplace transform, since we are interested in the slow, secular evolution of $f_0$ over a timescale much longer than the damping rate of the Landau modes.}, follows

\begin{align}
&\left<A_{\bk i}^\ast\left(\omega^{(\rmP)}\right) A_{\bk j}\left(\omega^{'(\rmP)}\right)\right> \nonumber\\
&= \calE_{ij}\left(\bk\right)\, \calC_{\omega}\left(\omega^{(\rmP)}\right) \delta\left(\omega^{(\rmP)}-\omega^{'(\rmP)}\right),
\label{white_noise_omega}
\end{align}
where $\calC_{\omega}$ is the Fourier transform of $\calC_t$.

Substituting the expressions for the linear quantities, $\bE_{\bk}(t)$ and $f_{1\bk}(\bv,t)$, from equations~(\ref{Ek_f1k_long_time_app}) in the quasilinear equation~(\ref{quasilin_resp_eq}) above, and using the noise spectrum for the perturbing electric field given in equation~(\ref{white_noise_t}), we obtain the following simplified form for the quasilinear equation (refer to Appendix~\ref{App:perturbation_theory} for a detailed derivation):

\begin{align}
\frac{\partial f_0}{\partial t} &= \frac{\partial}{\partial v_i}\left(D_{ij}(\bv)\frac{\partial f_0}{\partial v_j}\right).
\label{quasilin_resp_FP_eq}
\end{align}
This is nothing but a Fokker-Planck equation with a diffusion tensor $D_{ij}$, which at long time after the Landau modes have damped away (assuming that we are always in the stable regime), is given by

\begin{align}
D_{ij}(\bv) &\approx \frac{2^d \pi^{d+1} e^2}{m^2 V} \int \rmd^d k\, \frac{\calE_{ij}(\bk)\, \calC_{\omega}\left(\bk\cdot\bv\right)}{{\left|\varepsilon_{\bk}\left(\bk\cdot\bv\right)\right|}^2},
\label{diffusion_tensor}
\end{align}
with the dielectric constant, $\varepsilon_{\bk}\left(\bk\cdot\bv\right)$, given by the third of equations~(\ref{lin_resp_eq}). For perturbation on super-Debye scales ($k\lambda_\rmD \ll 1$), which is what we shall focus on throughout the paper, the Landau modes damp away at a rate faster than the quasilinear relaxation rate by a factor of $\sim {\left(k\lambda_\rmD\right)}^{-2}$. Hence, we neglect the decaying Landau term in the above derivation of the quasilinear diffusion tensor. The quasilinear diffusion equation~(\ref{quasilin_resp_FP_eq}) is also known as the secular dressed diffusion equation \citep[][]{Chavanis.23}, and has been previously derived by \citet[][]{Chavanis.12,Chavanis.22,Chavanis.23} in the astrophysical context.

The quasilinear diffusion of $f_0$ is governed by the fluctuating background as well as the self-consistent electric field generated by the fluctuating particles themselves. In fact, the diffusion is driven by the {\it polarized} or {\it dressed} fluctuations. Even though collisionless relaxation is in general a complicated, violent and turbulent process, the long-time relaxation of the coarse-grained mean DF of the bulk plasma is governed by a surprisingly simple Fokker-Planck type diffusion. In the quasi-steady state, $f_0$ becomes a particular function of velocity or energy, which appears to depend on the exact functional form of the diffusion tensor. However, we shall show below that, under a wide range of circumstances, the diffusion coefficient has a unique $v$ dependence over a large range of $v$, which ultimately leads to a universal $v$ dependence of $f_0$ in the quasi-steady state.

\section{The quasilinear diffusion coefficient}\label{sec:QL_diff_coeff}

\begin{figure}[t!]
\centering
\includegraphics[width=1\textwidth]{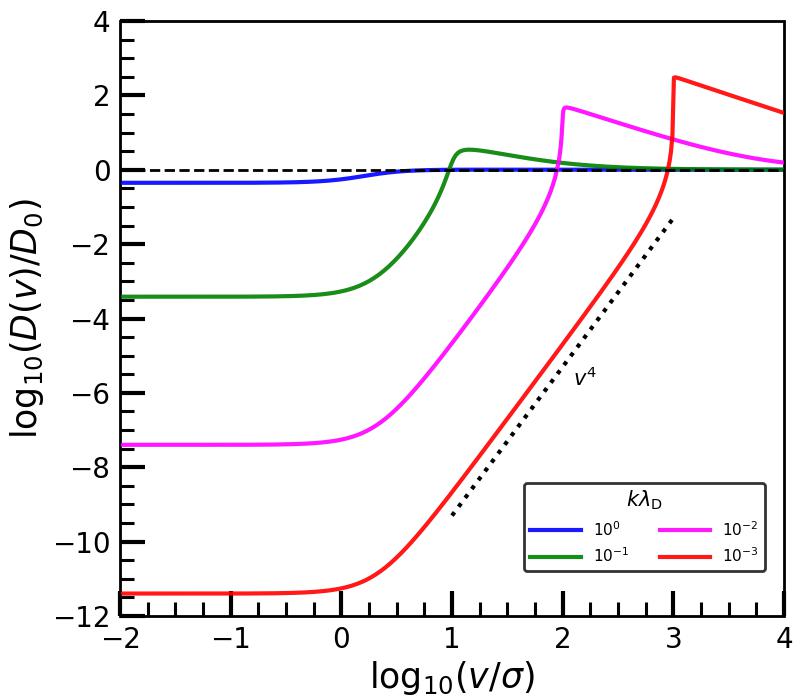}
\caption{The quasilinear diffusion coefficient, $D(v)$, normalized by $D_0 = 32\pi^5 e^2 k^2 \calE_0 /m^2 V$, as a function of $v/\sigma$ for a white noise forcing (with a single wavenumber $k$ such that the power spectrum is $\calE(k') = \calE_0\delta\left(k'-k\right)$) of a collisionless plasma characterized by $f_0$ that follows a $\kappa$ distribution with $\kappa=1$ and velocity dispersion, $\sigma$. Different lines indicate different values of $k\lambda_\rmD$, where $\lambda_\rmD=\sigma/\omega_\rmP$ is the Debye length. Note that $D(v)\sim v^4$ for $\sigma \lesssim v \lesssim \omega_\rmP/k = \sigma/k\lambda_\rmD$. This range widens as $k\lambda_\rmD$ decreases, i.e., for larger scale forcing. For smaller values of $k\lambda_\rmD$, $D(v)$ spikes at $v=\omega_\rmP/k$ since these particles are resonant with the plasma waves (electron Langmuir waves or ion acoustic waves) and extract the maximum energy from the electric field.}
\label{fig:diff_coeff_wp1}
\end{figure}

\begin{figure}[t!]
\centering
\includegraphics[width=1\textwidth]{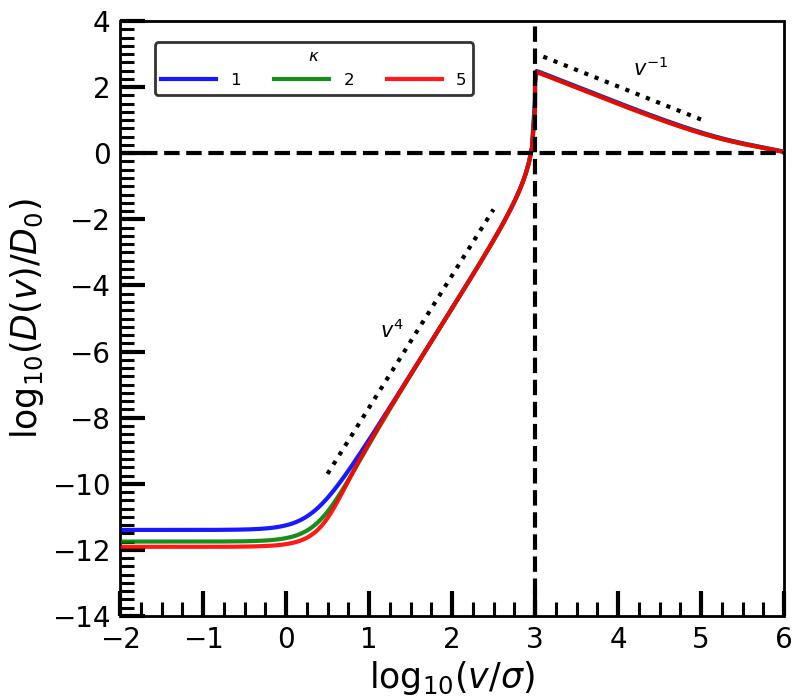}
\caption{Same as Fig.~\ref{fig:diff_coeff_wp1} but for different values of $\kappa$ as indicated and $k\lambda_\rmD = 10^{-3}$. Note that $D(v)$ is largely insensitive to $\kappa$ and scales as $v^4$ for $\sigma \lesssim v \lesssim \omega_\rmP/k = \sigma/k\lambda_\rmD = 10^3 \sigma$ regardless of $\kappa$.}
\label{fig:diff_coeff}
\end{figure}

As discussed above, the long time quasilinear evolution of the mean coarse-grained DF of a driven collisionless plasma is governed by a Fokker-Planck equation. The diffusion tensor, given by equation~(\ref{diffusion_tensor}), depends on the dielectric constant, which in turn depends on the DF through a velocity integral (see the third of equations~[\ref{lin_resp_eq}]). Therefore, in general, we have to numerically integrate an integro-differential equation in $d$ dimensions to track the temporal evolution of $f_0(\bv,t)$. We can, however, reduce the dimensionality of the problem by making the following simplifying assumptions:

\begin{itemize}
    \item Isotropic turbulence of the drive: $\calE_{ij}(\bk) = \calE(k)\, \delta_{ij}$
    \item Isotropic DF: $f_0(\bv) = f_0(v)$.
\end{itemize}
Under these assumptions, the quasilinear diffusion equation~(\ref{quasilin_resp_FP_eq}) is simplified into the following one-dimensional diffusion equation:

\begin{align}
&\frac{\partial f_0}{\partial t} = \frac{1}{v^{d-1}}\frac{\partial}{\partial v}\left(D(v)\, v^{d-1} \frac{\partial f_0}{\partial v}\right),
\label{QL_eq}
\end{align}
with the diffusion coefficient given by

\begin{align}
&D(v) = \frac{2^d \pi^{d+1} e^2}{m^2 V} \int_0^{\infty} \rmd k\, k^{d-1} \calE(k) \int \rmd\Omega_{\rmd}\, \frac{ \calC_{\omega}\left(k v \cos{\theta}\right)}{{\left|\varepsilon_{k}\left(kv\cos\theta\right)\right|}^2},
\end{align}
where $d$ is the number of dimensions, $\theta$ is the angle between $\bk$ and $\bv$, and $\rmd\Omega_\rmd$ is the differential solid angle in $d$ dimensions (equal to $\rmd \cos{\theta}$ in 3D). The dielectric constant is given by

\begin{align}
&\varepsilon_{k}\left(kv\cos\theta\right) = 1 + \frac{\omega^2_\rmP}{k^2} \int \rmd v'\, \frac{{\partial F_0}/{\partial v'}}{v\cos\theta - v'}\nonumber\\
&= 1 + \frac{\omega^2_\rmP}{k^2} \left[\int \rmd v'\,\frac{\partial F_0}{\partial v'}\, {\rmP} \left(\frac{1}{v\cos\theta - v'}\right) - i\pi \left.\frac{\partial F_0}{\partial v}\right|_{v\cos\theta}\right],
\end{align}
with $\rmP$ denoting the principal value, and

\begin{align}
F_0(v_x) = \prod_{i=2}^{d} \int \rmd v_i\, f_0(\bv)
\end{align}
the one-dimensional or reduced DF, $v_x$ being the component of $\bv$ along $\bk$. Clearly, the solution depends on the form of $D(v)$, which in turn depends on the dielectric constant, $\varepsilon_{k}$, and the temporal correlation, $\calC_t$, of the external perturbations. In what follows, we compute $D(v)$ for some physically well- motivated models of $\calC_t$ and study its asymptotic scalings. We shall hereafter assume $d=3$ for the analysis of $D(v)$, but we have checked that the velocity scaling of $D(v)$ does not depend on the number of dimensions.

\subsection{White noise}\label{sec:white_noise}

Let the external drive be a white noise, or in other words uncorrelated in time, in which case the correlation function, $\calC_t$, is of the following form:

\begin{align}
\calC_t\left(t,t'\right) = \delta\left(t-t'\right).
\label{Ct_white_noise}
\end{align}
Although idealized, this model is valid as long as the correlation time of the noise is shorter than the relevant dynamical timescale of the system, which is the plasma oscillation period, $2\pi/\omega_\rmP$. We discuss the implications of finite correlation time or redness of the noise in section~\ref{sec:red_noise} by adopting a specific example of $\calC_t$.

In the case of the white noise, where $\calC_t$ is given by equation~(\ref{Ct_white_noise}), we have $\calC_{\omega}\left(\bk\cdot\bv\right)=1$, and the diffusion coefficient simplifies to the following:

\begin{align}
&D(v) = \frac{32\pi^5 e^2}{m^2 V} \int_0^{\infty} \rmd k\, k^2 \calE(k) \int_{0}^{1}\rmd\left(\cos{\theta}\right) \frac{1}{{\left|\varepsilon_{k}\left(kv\cos\theta\right)\right|}^2}.
\label{diff_coeff}
\end{align}
As shown in Appendix~\ref{App:diff_coeff}, we can approximate $\varepsilon_k$ as $1 - \omega^2_\rmP/k^2 v^2 \cos^2{\theta}$ for $\sigma\lesssim v \lesssim \omega_\rmP/k$, and as $1 + c_{\rmF}\,\omega^2_\rmP/k^2\sigma^2$ for $v \lesssim \sigma$ to obtain an approximate analytical expression for $D(v)$ (see equation~[\ref{D_approx_wn_app}]). Here $c_\rmF$ is an $\calO(1)$ constant that depends on the high $v$ asymptotic behavior of $F_0$. The following asymptotic scalings of $D(v)$ are important:

\begin{widetext}
\begin{align}
&D(v) \approx \dfrac{32\pi^5 e^2}{m^2 V} \times 
\begin{cases}
\bigintsss_{\;0}^{\infty} \rmd k\, \frac{\mathlarger{k^2 \calE(k)}}{{\left(\mathlarger{1 + c_{\rmF}} \dfrac{\omega^2_\rmP}{k^2\sigma^2}\right)}^{\mathlarger{2}}}, \qquad \qquad v \ll \sigma, \\
\dfrac{v^4}{5 \omega^4_\rmP}\, \bigintsss_{\;0}^{\infty} \rmd k\, k^6 \calE(k), \qquad \qquad \quad \sigma \ll v \ll \dfrac{\omega_\rmP}{k}, \\ \\
\dfrac{\omega_\rmP}{v} \,\bigintsss_{\;0}^{\infty} \rmd k\, k\, \calE(k), \qquad \qquad \quad \;\, \dfrac{\omega_\rmP}{k} \ll v \ll \dfrac{1}{k\lambda_\rmD}\dfrac{\omega_\rmP}{k}, \\ \\
\bigintsss_{\;0}^{\infty} \rmd k\, k^2 \calE(k), \qquad \qquad \qquad \quad\, v \gg \dfrac{1}{k\lambda_\rmD}\dfrac{\omega_\rmP}{k}.
\end{cases}
\label{diff_coeff_asymptotic_wn}
\end{align}
\end{widetext}
Although the above scalings are asymptotic, they hold even for velocities close to the limits, as one can see from Fig.~\ref{fig:diff_coeff_wp1} that plots $D(v)$ as a function of $v$ for white noise forcing. Thus we shall hereafter use the $\lesssim$ and $\gtrsim$ symbols to specify the velocity ranges with different asymptotic scalings. When $f_0(v)$ is a $\kappa$ distribution, i.e.,

\begin{align}
f_0(v) = \frac{1}{{\left(2\pi\sigma^2\right)}^{3/2}} \frac{\Gamma\left(\kappa+1\right)}{\kappa^{3/2}\Gamma\left(\kappa-1/2\right)} \frac{1}{{\left(1+\dfrac{v^2}{2 \kappa \sigma^2}\right)}^{1+\kappa}},
\label{kappa_dist}
\end{align}
the constant $c_{\rmF}$ is equal to $\left(1-1/2\kappa\right)$.


Assuming a spatially sinusoidal drive, i.e., $\calE(k')=\calE_0\,\delta\left(k'-k\right)$, we numerically compute the diffusion coefficient given in equation~(\ref{diff_coeff}) for a $\kappa$ distribution with $\kappa=1$ (as we show below, the result is not sensitive to the value of $\kappa$). We plot the $D(v)$ thus obtained, normalized by $D_0 = 32\pi^5 e^2 k^2 \calE_0 /m^2 V$, in Fig.~\ref{fig:diff_coeff_wp1} as a function of $v$ for $k\lambda_\rmD=1$, $10^{-1}$, $10^{-2}$ and $10^{-3}$. On scales comparable to the Debye length, there is no Debye shielding and therefore $\left|\varepsilon_{\bk}\right| \approx 1$, implying $D(v)\approx D_0$. On larger scales, due to Debye screening of the electric field, $\left|\varepsilon_{\bk}\right|>1$ and thus, $D(v)<D_0$ for $v \lesssim \omega_\rmP/k$, the phase-velocity of the plasma waves. In the $v\ll \sigma$ limit, $D(v) \approx D_0 {\left(1-1/2\kappa\right)}^{-2} {\left(k\lambda_\rmD\right)}^4$, and then increases with $v$ as $\sim v^4$ from $v\sim \sigma$ up to $v\sim\omega_\rmP/k$. It sharply increases as $v\to \omega_\rmP/k$ due to wave-particle resonance, and decreases thereafter as $v^{-1}$ until it saturates to $D_0$ at $v \gg \omega_\rmP/k$. The velocity range over which $D(v)$ scales as $v^4$ increases linearly with ${\left(k\lambda_\rmD\right)}^{-1}$, i.e., widens for larger scale forcing. Particles moving slower than the phase-velocity of the plasma waves experience a large-scale electric field that is Debye screened, while those moving faster experience the bare field. This is the reason why the faster particles carry more effective charge than the slower ones, and are consequently heated more than the latter. Those moving with $v\approx \omega_\rmP/k$ resonate with the plasma waves, thereby extracting maximum energy from the electric field. This wave-particle resonance leads to the sharp increase of the diffusion coefficient near $v=\omega_\rmP/k$. The non-resonant particles undergo much less heating, their diffusion being suppressed by a factor of $\approx {\left(k v /\omega_\rmP\right)}^{-4}$ relative to that of the resonant particles; this scaling can be traced fundamentally to the inverse square nature of the Coulomb force. The core of the DF, consisting of particles with $v \lesssim \sigma$, diffuses very little, at a rate suppressed by a factor of $\approx {\left(k\lambda_\rmD\right)}^{-4}$ with respect to the high energy particles. It is the suppression of the diffusion of the non-resonant particles relative to the resonant ones by a factor of ${\left(k v /\omega_\rmP\right)}^{-4}$ that ultimately gives rise to the universal velocity scaling of the quasi-steady state DF, as we shall see in the next section.

The velocity dependence of the diffusion coefficient is quite insensitive to the exact functional form of the DF for $v\gtrsim \sigma$. To demonstrate this, we plot $D(v)/D_0$ in Fig.~\ref{fig:diff_coeff} as a function of $v$ for the $\kappa$ distribution with different values of $\kappa$ as indicated. We adopt large-scale forcing, i.e., $k\lambda_\rmD=10^{-3}$. Note that $D(v)$ scales as ${\left(1-1/2\kappa\right)}^{-2}$ for $v\lesssim \sigma$, i.e., it weakens only slightly with increasing $\kappa$ or steeper large $v$ fall-off of the DF. Leaving aside this slight modification, $D(v)$ is largely insensitive to $\kappa$ elsewhere, and scales as $\sim v^4$ for $\sigma\lesssim v\lesssim \omega_\rmP/k$. This universal behavior of the quasilinear diffusion coefficient at intermediate velocities appears as long as the external forcing is acting isotropically on scales larger than the Debye length as a white noise in time.

\begin{figure}[t!]
\centering
\includegraphics[width=1\textwidth]{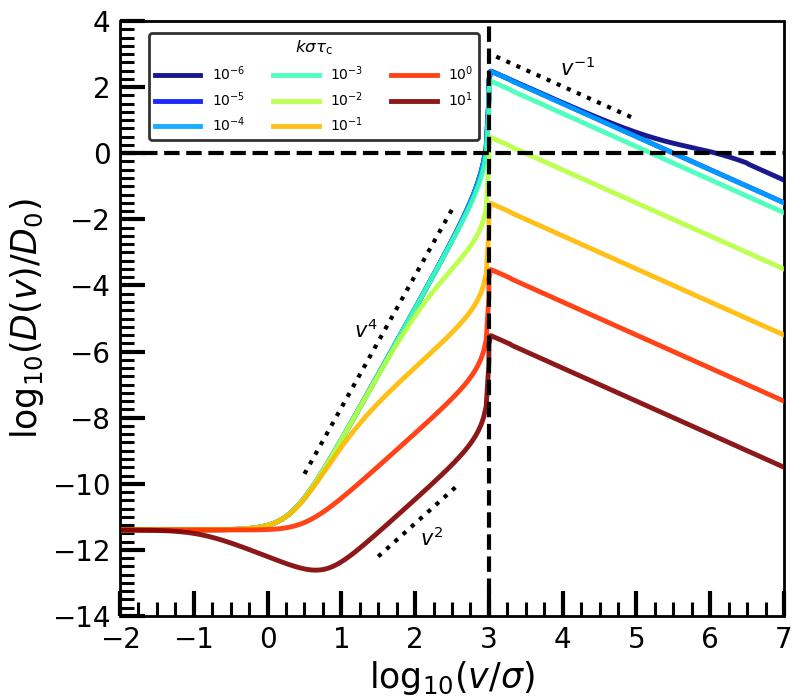}
\caption{Same as Fig.~\ref{fig:diff_coeff_wp1} but for a red noise drive (of the model given by equation~[\ref{red_noise_model}]) with different correlation times, $\tau_\rmc$, as indicated. Note how $D(v)$ scales the same way as in the white noise case, i.e., $\sim v^4$ for $\sigma \lesssim v \lesssim \omega_\rmP/k$, when $\omega_\rmP \tau_\rmc \lesssim 1$. For larger correlation times such that $\omega_\rmP\tau_\rmc \gtrsim 1 \gtrsim k\sigma\tau_\rmc$, the high velocity end between $v\sim 1/k\tau_\rmc$ and $\omega_\rmP/k$ develops a $v^2$ scaling, while for $\sigma \lesssim v \lesssim 1/k\tau_\rmc$ $D(v)$ still scales as $v^4$. The very high velocity end ($v\gtrsim \omega_\rmP/k$) develops a $v^{-1}$ scaling.}
\label{fig:diff_coeff_tc}
\end{figure}

\subsection{Red noise}\label{sec:red_noise}

The universal $v^4$ scaling of the diffusion coefficient is partially broken if we have red noise, i.e., a finite correlation time for the external electric field. Let us take a specific example of red noise to see this effect. Let the correlation function, $\calC_t$, be of the form:

\begin{align}
\calC_t\left(t-t'\right) = \frac{1}{2 \tau_\rmc} \exp{\left[-\left|t-t'\right|/\tau_\rmc\right]}.
\label{red_noise_model}
\end{align}
In this case,

\begin{align}
\calC_{\omega}\left(\bk\cdot\bv\right) = \frac{1}{1 + {\left(\bk\cdot\bv \, \tau_\rmc\right)}^2},
\end{align}
which tends to $1$ as $\tau_\rmc \to 0$, as one would expect since $\calC_t\left(t-t'\right) \to \delta\left(t-t'\right)$ and the red noise becomes white noise in this limit. 

With the above form for the noise, the quasilinear diffusion coefficient becomes

\begin{align}
D(v) &= \frac{32\pi^5 e^2}{m^2 V} \int_0^{\infty} \rmd k\, k^2 \calE(k) \nonumber\\
&\times \int_{0}^{1}\rmd\left(\cos{\theta}\right) \frac{1}{1 + {\left(k v \tau_\rmc \cos{\theta}\right)}^2} \frac{1}{{\left|\varepsilon_{k}\left(kv\cos\theta\right)\right|}^2}.
\end{align}
As shown in Appendix~\ref{App:diff_coeff_rn}, we can use an approximate expression for $\varepsilon_k$ and thus approximately evaluate an analytical expression for $D(v)$ (see equations~[\ref{D_approx_rn1_app}] and [\ref{D_approx_rn2_app}]). For $\omega_\rmP \tau_\rmc > 1 \gtrsim k\sigma \tau_\rmc$, $D(v)$ has the following asymptotic behavior:

\begin{widetext}
\begin{align}
D(v) &\approx \dfrac{32\pi^5 e^2}{m^2 V} \times
\begin{cases}
\bigintsss_{\;0}^{\infty} \rmd k\, \frac{\mathlarger{k^2 \calE(k)}}{{\left(\mathlarger{1 + c_{\rmF}} \dfrac{\omega^2_\rmP}{k^2\sigma^2}\right)}^{\mathlarger{2}}}, \qquad \qquad v \ll \sigma, \\
\dfrac{v^4}{5 \omega^4_\rmP} \bigintsss_{\;0}^{\infty}\rmd k\, k^6\, \calE(k), \qquad \qquad \quad \sigma \ll v \ll \dfrac{1}{k\tau_\rmc},\\ \\
\dfrac{4 v^2}{3 \omega^4_\rmP \tau^2_\rmc} \bigintsss_{\;0}^{\infty}\rmd k\, k^4\, \calE(k), \qquad \quad \;\;\; \dfrac{1}{k\tau_\rmc} \ll v \ll \dfrac{\omega_\rmP}{k},\\ \\
\dfrac{1}{\omega_\rmP \tau^2_\rmc v} \bigintsss_{\;0}^{\infty} \rmd k\, k\, \calE(k), \qquad \qquad \;\; v \gg \dfrac{\omega_\rmP}{k}.
\end{cases}
\label{diff_coeff_asymptotic_rn}
\end{align}
\end{widetext}
For $\omega_\rmP \tau_\rmc > k\sigma \tau_\rmc \gtrsim 1$, $D(v)$ is independent of $v$ for $v\lesssim 1/k\tau_\rmc$, scales as $v^{-1}$ for $1/k\tau_\rmc \lesssim v \lesssim \sigma$, $v^2$ for $\sigma \lesssim v \lesssim \omega_\rmP/k$ and $v^{-1}$ beyond. For $\omega_\rmP \tau_\rmc \lesssim 1$, $D(v)$ behaves the same way as in the white noise case for $v \lesssim 1/k\tau_\rmc$, but for $v \gtrsim 1/k\tau_\rmc$, becomes

\begin{figure}[t!]
\centering
\includegraphics[width=1\textwidth]{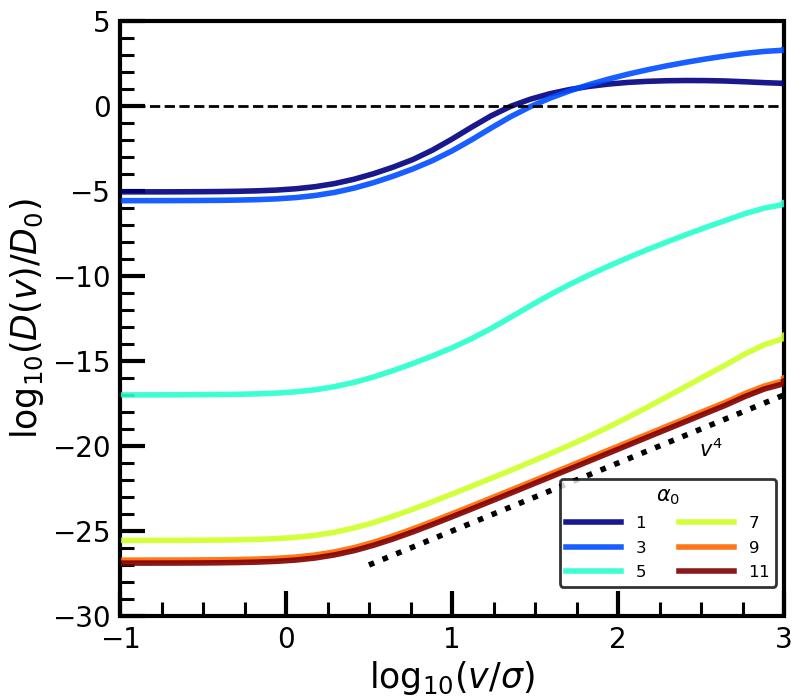}
\caption{Same as Fig.~\ref{fig:diff_coeff_wp1} but for $\calE(k)$ a Schechter function, given by equation~(\ref{Schechter_func}), with $k_\rmc \lambda_\rmD=10^{-2}$ and different values of $\alpha_0$ as indicated. $D(v)$ develops a more pronounced and more extended $v^4$ scaling (for $\sigma \lesssim v \lesssim \omega_\rmP/k_{\ast}$ with $k_\ast$ a characteristic wavenumber) as $\alpha_0$ increases, i.e., larger scales dominate.}
\label{fig:diff_coeff_alpha}
\end{figure}

\begin{figure}[t!]
\centering
\includegraphics[width=1\textwidth]{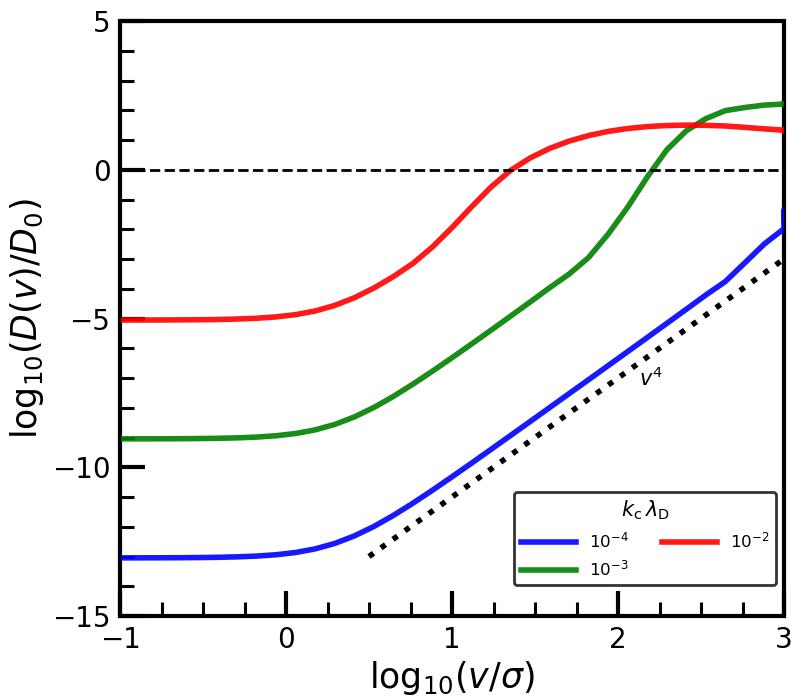}
\caption{Same as Fig.~\ref{fig:diff_coeff_wp1} but for $\calE(k)$ a Schechter function, given by equation~(\ref{Schechter_func}), with $\alpha_0=1$ and different values of the cut-off wavenumber $k_\rmc$ as indicated. $D(v)$ develops a more pronounced $v^4$ scaling (for $\sigma \lesssim v \lesssim \omega_\rmP/k_\rmc$) as $k_\rmc \lambda_\rmD$ decreases, i.e., larger scales dominate.}
\label{fig:diff_coeff_kc}
\end{figure}

\begin{align}
D(v) \approx \frac{16 \pi^6 e^2}{m^2 V \tau_\rmc} \frac{1}{v} \int_0^{\infty} \rmd k\, k\, \calE(k).
\end{align}

The above scalings are manifest in Fig.~\ref{fig:diff_coeff_tc} that plots $D(v)/D_0$ vs $v$ for different values of $k\sigma \tau_\rmc$ as indicated, adopting $k\lambda_\rmD=10^{-3}$. Note that in the $\omega_\rmP \tau_\rmc \to 0$ limit, we recover the same scalings as in the white noise case (notably, $D(v)\sim v^4$ over a large range in $v$) except for $v>1/k\tau_\rmc$. Therefore we see that a non-zero $\tau_\rmc$ does not destroy the universal $v^4$ scaling of the diffusion coefficient if $\tau_\rmc$ is shorter than the plasma oscillation period, $1/\omega_\rmP$. Strongly correlated noise (of the form given in equation~[\ref{red_noise_model}]) with $\tau_\rmc>1/\omega_\rmP$, on the other hand, can modify the velocity scaling to $\sim v^2$ for $v > 1/k\tau_\rmc$, but keeps the $v^4$ scaling unchanged for $\sigma < v < 1/k\tau_\rmc$.

\subsection{Spatial power spectrum of the external perturbation}

The diffusion coefficient depends on $\calE(k)$, the spatial power spectrum of the perturbation. To examine how strong this dependence is, we adopt a Schechter function for the power spectrum,

\begin{align}
\calE(k) = \calE_0\,\frac{3-\alpha_0}{1-{\left(k_{\rm min}/k_\rmc\right)}^{3-\alpha_0}}\, {\left(\frac{k}{k_\rmc}\right)}^{-\alpha_0} \exp{\left[-k/k_\rmc\right]}\,\Theta\left(k-k_{\rm min}
\label{Schechter_func}
\right)\end{align}
and substitute it in equation~(\ref{diff_coeff}) to compute the diffusion coefficient, adopting a $\kappa$ distribution with $\kappa=1$ (as we pointed out earlier, the result is insensitive to $\kappa$), and a white noise temporal power spectrum for the external perturbation. We plot the resulting $D(v)/D_0$ as a function of $v$ in Fig.~\ref{fig:diff_coeff_alpha} for $\alpha_0=1,3,5,7,9$ and $11$, fixing $k_\rmc = 10^{-2}$, and in Fig.~\ref{fig:diff_coeff_kc} for $k_\rmc = 10^{-2},10^{-3}$ and $10^{-4}$, fixing $\alpha_0=1$. We adopt $k_{\rm min}\lambda_\rmD=10^{-7}$. Note that $D(v)\sim v^4$ in the range $\sigma<v\lesssim \omega_\rmP/k_{\ast}$, with $k_{\ast}\approx k_\rmc$ for $\alpha_0 \ll 7$ and $k_{\ast}\approx k_{\rm min}$ for $\alpha_0\gg 7$. This can be understood from the $k$ integral in the expression for $D(v)$ in equation~(\ref{diff_coeff_asymptotic_wn}) when $D(v)\sim v^4$. Smaller values of $\alpha_0$ constrain the $v^4$ scaling of $D(v)$ to a smaller range in $v$. The $v^4$ scaling persists for a larger range of $v$ for larger scale forcing, i.e., for smaller values of $k_\rmc \lambda_\rmD$ and/or $k_{\rm min}\lambda_\rmD$ as well as for larger values of $\alpha_0$. The range of $v^4$ scaling is sensitive to $k_\rmc$ ($k_{\rm min}$) and insensitive to $k_{\rm min}$ ($k_\rmc$) for small (large) $\alpha_0$.
\\

\section{The quasilinear distribution function}\label{sec:QL_DF}

\begin{figure*}
\centering
\includegraphics[width=0.7\textwidth]{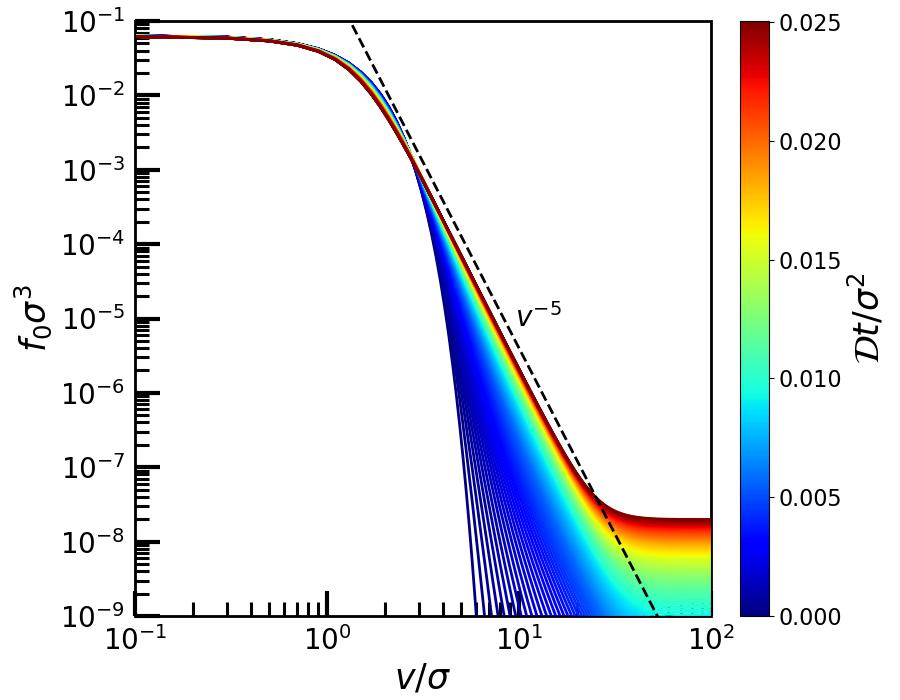}
\caption{Quasilinear evolution of the mean coarse-grained DF, $f_0(v)$, as a function of $v$, obtained by solving equation~(\ref{QL_diffusion_eqn_nd}) in 3D (see section~\ref{sec:QL_DF} for details), starting from an initial Maxwellian. The system is driven by a white noise field of a single wavenumber $k=10^{-2}/\lambda_\rmD$. Different colors denote different times in units of $\sigma^2/\calD$ with $\calD = {\left(k\lambda_\rmD\right)}^4 D_0$ and $D_0 = 32\pi^5 e^2 k^2 \calE_0 /m^2 V$. Note how a $v^{-5}$ power-law tail develops for $\sigma \lesssim v \lesssim \omega_\rmP/k = 100\sigma$ with $\sigma$ the velocity dispersion of the initial DF. The high velocity end ($v\gtrsim \omega_\rmP/k$) forms a plateau. The Maxwellian core ($v\lesssim \sigma$) heats up over a much longer timescale $\sim {\left(k\lambda_\rmD\right)}^{-4} \sigma^2/2 D_0$ than that over which the power-law tail forms.}
\label{fig:f_evol}
\end{figure*}

\begin{figure*}
\centering
\includegraphics[width=0.7\textwidth]{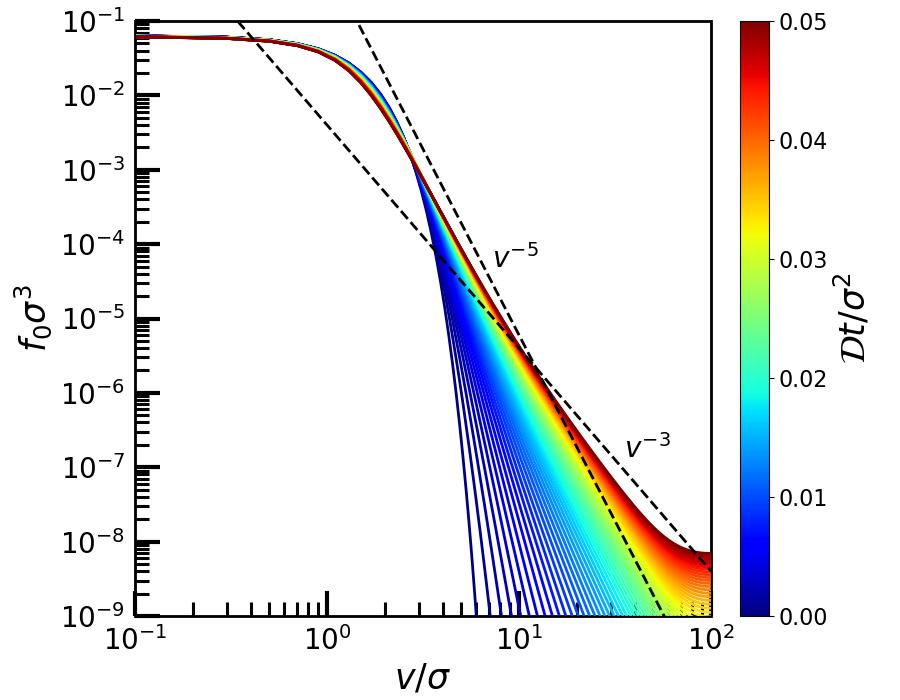}
\caption{Same as Fig.~\ref{fig:f_evol} but for a red noise drive of the form given in equation~(\ref{red_noise_model}) with a correlation time $\tau_\rmc = 10/\omega_\rmP$. Note that a $v^{-5}$ power-law tail still forms but for $\sigma \lesssim v \lesssim 1/k\tau_\rmc = 10\sigma$, while a shallower $\sim v^{-3}$ tail forms for $10\sigma = 1/k\tau_\rmc \lesssim v \lesssim \omega_\rmP/k = 100\sigma$. To compensate for the formation of this harder spectrum, the high velocity ($v\gtrsim \omega_\rmP/k$) plateau forms slower than in the white noise case.}
\label{fig:f_evol_corr}
\end{figure*}

We numerically solve the quasilinear diffusion equation~(\ref{QL_eq}) using the diffusion coefficient evaluated in the previous section. We use a finite difference flux-conserving scheme, outlined in Appendix~\ref{App:code}, to solve the diffusion equation, which, in the non-dimensional form, can be written as: 

\begin{align}
\frac{\partial f_0}{\partial \tau} = \frac{1}{u^{d-1}}\frac{\partial}{\partial u}\left(\Tilde{D}(u)\, u^{d - 1} \frac{\partial f_0}{\partial u}\right),
\label{QL_diffusion_eqn_nd}
\end{align}
with $u=v/\sigma$, $\Tilde{D}(u) = D(u)/\calD$, $\calD = D_0{\left(k\lambda_\rmD\right)}^4$, $\tau = \calD t/\sigma^2$, $D_0 = D(v\ll\sigma) = 32\pi^5 e^2 k^2 \calE_0 /m^2 V$, and $\sigma$ the velocity dispersion of the initial DF. For simplicity, we assume the power spectrum of the external drive to have non-zero contribution from a single super-Debye $k$ mode ($k\lambda_\rmD \ll 1$). For a more complicated $\calE(k)$ such as the Schechter function given in equation~(\ref{Schechter_func}), the results are unchanged if the forcing is dominated by the largest scales ($k_\rmc \lambda_\rmD \ll 1$ or $\alpha_0\gtrsim 7$). As boundary condition, we assume that the flux $\propto \partial f_0/\partial u$ is zero at the boundaries in $u$. 

In Fig.~\ref{fig:f_evol}, we plot the $f_0$ of a plasma in 3D, driven by a white noise ($\omega_\rmP \tau_\rmc \lesssim 1$) electric field with $k\lambda_\rmD = 10^{-2}$, as a function of $v$, for different times, starting from a Maxwellian distribution. The DF rapidly develops a power-law tail, and the power-law exponent asymptotes to $-5$, i.e., $f_0$ assumes the form,

\begin{align}
f_0(v)\sim v^{-5},
\end{align}
due to the $v^4$ dependence of $D(v)$ in the range, $\sigma < v \lesssim \omega_\rmP/k = \sigma/k\lambda_\rmD = 100\sigma$. In fact, the DF looks very much like the $\kappa$ distribution (equation~[\ref{kappa_dist}]) with $\kappa=1.5$ for $v\lesssim \omega_\rmP/k$. Meanwhile, the high $v$ end of the DF develops a quasilinear plateau quite rapidly over a timescale $\sigma^2/2 D_0$, which indicates the escape of high velocity particles with $v\sim \omega_\rmP/k$ due to near-resonant wave-particle interactions. We plot the evolution of $f_0$ for red noise forcing with $\omega_\rmP \tau_\rmc = 10$ in Fig.~\ref{fig:f_evol_corr}. This differs from the white noise case in the fact that, in the red noise case, the $v^{-5}$ tail appears over a more restricted range ($\sigma < v \lesssim 1/k\tau_\rmc = 10\sigma$) while the power-law tail assumes a shallower $\sim v^{-3}$ form for $v\gtrsim 1/k\tau_\rmc = 10\sigma$ (since $D(v)\sim v^2$ for $1/k\tau_\rmc \lesssim v < \omega_\rmP/k = 100\sigma$). In the $\sigma \lesssim v \lesssim 1/k\tau_\rmc$ range, the DF develops the same $v^{-5}$ power-law tail for both white and red noise forcing, since these particles move slower than the rate at which the external field decorrelates and therefore experience this stochastic field as essentially a white noise. For $1/k\tau_\rmc \lesssim v \lesssim \omega_\rmP/k$, however, the particles experience a similar strength of the field for a longer time since they move faster than the rate of decorrelation of the field. This implies a weaker velocity dependence of $D(v)$ for $v \gtrsim 1/k\tau_\rmc$ (see Fig.~\ref{fig:diff_coeff_tc}) and a more comparable heating of these particles, thereby leading to a harder spectrum. 

The above scalings of the DF can be understood through an approximate self-similar solution of the diffusion equation assuming a power-law form for $D(v)$ (it behaves as a power-law with a different exponent in a different velocity interval, as discussed in section~\ref{sec:QL_diff_coeff}), as detailed in Appendix~\ref{App:self_similar_sol}. The $v^{-5}$ and $v^{-3}$ forms for $f_0$ are also the quasi-steady state solutions of the diffusion equation~(\ref{QL_diffusion_eqn_nd}), obtained by assuming $D(v)\sim v^4$ and $v^2$ respectively and taking $\partial f_0/\partial t = 0$. In $d$ dimensions, the above power-law scalings of $f_0$ become $v^{-\left(d+2\right)}$ and $v^{-d}$ respectively.

The DF remains a Maxwellian for $v\lesssim \sigma$. The Maxwellian core can naturally arise from weak collisions. Although we consider collisionless relaxation in this paper, no plasma in nature is truly collisionless, rather they harbor weak collisions that tend to Maxwellianize the core of the distribution over a long time \citep[][]{Banik.Bhattacharjee.24}. However, if the plasma is subject to external heating, the DF would naturally develop a power-law tail due to the power-law $v$ dependence of the diffusion coefficient. As discussed in section~\ref{sec:QL_diff_coeff}, this arises from the power-law suppression of the diffusion of the slower particles relative to the faster ones due to the enhanced Debye screening of the electric field experienced by the former. The temperature of the Maxwellian core also increases, albeit gradually, over a very long period, 

\begin{align}
t_{\rm relax} \approx \frac{1}{{\left(k\lambda_\rmD\right)}^4} \frac{\sigma^2}{2 D_0},
\end{align}
since $k\lambda_\rmD \ll 1$. This `relaxation' timescale is far longer than the timescale over which the power-law tail develops. It is also much longer than the damping time of the Landau modes, $t_{\rm Landau} \approx {\left(k\lambda_\rmD\right)}^{-2}\left(2/3\pi \omega_\rmP\right)$ (assuming $f_0(v)\sim v^{-5}$), which implies that the Landau modes damp away faster than the rate of quasilinear relaxation, justifying our assumption of the Landau term being negligible in the derivation of the quasilinear diffusion tensor (equation~[\ref{diffusion_tensor}]). Weak collisions would typically only Maxwellianize the core of the distribution but not affect the power-law tail. Based on the above, we conclude that the non-thermal power-law tail thrives forever in a collisionless (or weakly collisional) plasma, as long as the system continues to undergo stochastic forcing on scales much larger than the Debye length. If this stochastic drive is spatially isotropic and weakly correlated in time, i.e., white noise-like, then the non-thermal tail has a universal $v^{-5}$ scaling.

We would like to draw the attention of the reader to the fact that $v^{-5}$ is the universal scaling of the DF only in 3D. In $d$ dimensions, we have the following scaling:

\begin{align}
f_0(v) \sim v^{-\left(d+2\right)},
\end{align}
i.e., it scales as $v^{-3}$ in 1D, $v^{-4}$ in 2D, and $v^{-5}$ in 3D. Recalling that the density of states, $g(v)$, scales as

\begin{align}
g(v) \sim v^{d-2},
\end{align}
we have the following scaling for the energy distribution or the number of particles per unit energy:

\begin{align}
N(E) = g(E) f(E) \sim E^{-2},
\end{align}
where $E=v^2/2$. This scaling is independent of dimensionality and is, therefore, a general result.


\section{Discussion and summary}\label{sec:discussion_summary}

In this paper, we study the quasilinear relaxation of a driven collisionless electrostatic plasma, and the evolution of the mean coarse-grained distribution function, $f_0$, in the process. Curiously, we discover that the quasi-steady state $f_0$ self-similarly scales as $v^{-\left(d+2\right)}$ ($d$ is the number of dimensions), or equivalently the energy distribution scales as $E^{-2}$, over a large range in $v$, irrespective of the initial conditions, as long as the following conditions are met:

\begin{itemize}
    \item The system is forced on scales larger than the Debye length. 

    \item The external electrostatic forcing is isotropic.

    \item The external forcing is white noise-like (small correlation time).
\end{itemize}

How universal is this $v$ dependence of $f_0$? Interestingly, apart from the condition of isotropy, $f_0$ has no dependence on the detailed spatial structure of the external perturbation, i.e., the exact power spectrum of the drive, $\calE(k)$. Typically, in a turbulent environment, $\calE(k)$ is a self-similar, power-law function of $k$ within the inertial range of the turbulent cascade (the inertial range spectrum $\calE(k)$ for Kolmogorov turbulence has a $k^{-11/3}$ dependence). However, the quasilinear diffusion coefficient, $D(v)$, scales as $v^4$, irrespective of the power-law exponent of the turbulent spectrum, as long as it predominantly acts on scales larger than $\lambda_\rmD$. When such large scale forcing occurs in a nearly uncorrelated fashion over time (white noise-like), $D(v)$ naturally develops a $v^4$ scaling and the corresponding $f_0$ scales as $v^{-5}$ (in 3D) over the velocity range, $\sigma < v \lesssim \omega_\rmP/k$, $k$ being the cut-off wavenumber of the drive. In this sense of insensitivity to the functional form of $\calE(k)$ and the initial condition, the $v^{-5}$ scaling of $f_0$ is universal.

The universality is partially broken by the violation of any of the aforementioned conditions for external forcing. This happens, for example, when the external drive is a red noise with a correlation time $\tau_\rmc$ such that $\omega_\rmP \tau_\rmc \gtrsim 1 \gtrsim k\sigma \tau_\rmc$. In this case, the $v^4$ scaling of the diffusion coefficient is untouched for all $\sigma < v < 1/k\tau_\rmc$, but significantly modified for $1/k\tau_\rmc < v < \omega_\rmP/k$. If the temporal correlation of the red noise is exponential in time, then the modified scaling of $D(v)$ turns out to be $v^2$. This implies that $f_0$ still scales as $v^{-5}$ in the range, $\sigma < v < 1/k\tau_\rmc$, but as roughly $v^{-3}$ for $1/k\tau_\rmc < v < \omega_\rmP/k$. If, on the other hand, $\omega_\rmP\tau_\rmc \lesssim 1$, then $D(v)$ scales the same way with $v$ as in the white noise case for $v<1/k\tau_\rmc$ but as $v^{-1}$ for larger velocities. In this case, only the high energy end of the distribution ($v>1/k\tau_\rmc$) is affected by the correlated nature of the noise. All in all, the $v^{-5}$ scaling of $f_0$ appears for sufficiently large-scale (isotropic) forcing with a sufficiently small correlation time.

What is the physics behind the $v^4$ scaling of the quasilinear diffusion coefficient and the consequent $v^{-5}$ scaling of the quasi-steady state DF? Slower particles tend to experience a larger scale perturbation since they take a longer time to traverse one full wavelength of the external field. Hence, they feel a weaker/more strongly Debye-screened field. This manifests as the `dressing' of the external field and the corresponding response by the dielectric constant, which scales as $\sim \omega^2_\rmP/k^2 v^2$ for $v \lesssim \omega_\rmP/k$ with $k$ a characteristic wavenumber of the perturbation power spectrum. This scaling ultimately arises from the inverse square law nature of the Coulomb force. Due to the dielectric polarization of the medium, slower particles are more Debye-screened and end up with a smaller effective charge, thereby undergoing less acceleration than the faster, less Debye-screened particles with higher effective charge. This leads to an uneven, velocity-dependent acceleration of the particles, and naturally pushes the high- energy end of an initially Maxwellian DF to higher energies, producing a power-law tail. As shown in section~\ref{sec:QL_DF}, this scaling turns out to be $v^{-5}$ in the velocity range, $\sigma<v<\omega_\rmP/k$, for a white noise drive, but can deviate from it in the high $v$ end for a red noise drive with a correlation time longer than the plasma oscillation period. The core of the Maxwellian ($v<\sigma$) is also heated, i.e., $\sigma$ increases, but over a much longer period that scales as $\sim {\left(k\lambda_\rmD\right)}^{-4}$, since the diffusion coefficient in the low-velocity end is suppressed by a factor of ${\left(k\lambda_\rmD\right)}^{-4}$ relative to the near-resonant particles ($v\sim \omega_\rmP/k$). The very high-velocity end, $v\gtrsim \omega_\rmP/k$, develops a plateau due to the efficient heating of the near-resonant particles.

A key requirement for the emergence of the $v^{-5}$ tail is the driving of the plasma on scales sufficiently larger than the Debye length. A large class of plasma waves, both electrostatic\footnote{In two-stream instabilities of electrostatic plasmas \citep[][]{Ewart.etal.24}, the bulk plasma is electrostatically driven by the BGK holes on super-Debye scales over long time. This may be the reason why the $f_0$ from our quasilinear formalism and that from their 1D PIC simulations both show an $E^{-2}$ tail.} and electromagnetic, satisfies this basic criterion. Here we have focused only on electrostatic perturbations. In the solar wind, it is widely known that the typical electric field spectrum is accounted for by the quasi-thermal noise of the electrons and the Doppler-shifted thermal fluctuations of the ions (see, for example, the monograph by Meyer-Vernet \citep[][]{Meyer-Vernet.07} for an excellent discussion). That being said, our calculation should be viewed as a prototypical application of QLT, one that can be generalized to cover a broad range of wave-particle interactions, with and without magnetic fields. The crucial point of principle in our calculation is the inclusion of self-consistency whereby the back-reaction of the fields generated by the charged particles on their DF is obtained by coupling to Maxwell's equations (in the electrostatic case, this simply reduces to the Poisson's equation). For the problem of ion acceleration in the solar wind, acceleration mechanisms such as stochastic acceleration have been considered by \citet[][]{Jokipii.Lee.10} by using the well-known Parker equation \citep[][]{Parker.65}, but without the constraints of self-consistency. In a separate publication, we will apply our methodology to the Parker equation, including self-consistency. We note that Jokipii and Lee's primary criticism of Fisk \& Gloecker is that the latter's transport equation does not conserve particle number, which does not apply to our transport equation~(\ref{quasilin_resp_FP_eq}) that explicitly conserves it.

We would like to emphasize here that the presence of large-scale electric fields is not uncommon in collisionless plasmas and is often associated with non-thermal particle acceleration. For example, the magnetic reconnection layer is known to harbor super-Debye electric fields \citep[][]{Hesse.etal.18} that are thought to be responsible for the formation of non-thermal heavy tailed distributions \citep[][]{Sironi.Spitkovsky.14,Hoshino.22}, similar to the $E^{-2}$ energy distribution we predict, albeit for relativistic electrostatic plasmas. Ground-based geomagnetic observations have confirmed that super-Debye fields are generated in the earth's magnetosphere by the interaction of the solar wind with the outer geomagnetic field and are responsible for the activation of auroral electrojets and current vortices \citep[][]{Obayashi.Nishida.68}. Collisionless shocks in astrophysical environments, e.g., supernova blast waves, often harbor super-Debye electric fields (e.g., sourced by the cross-shock potential) that can cause non-thermal acceleration of cosmic rays reflecting back and forth off the magnetic mirrors and repeatedly crossing the shock, which is the basic mechanism of Fermi acceleration \citep[][]{Fermi.49} and diffusive shock acceleration (DSA). PIC simulations of such collisionless shocks \citep[][]{Gupta.etal.24} find that the momentum ($p$) distribution of electrons and protons is often steeper than the $p^{-4}$ form predicted by the standard (non self-consistent) DSA theory for strong shocks. This might be because the cosmic rays and magnetic mirrors drift behind the shock (``postcursor"), something that self-consistent kinetic simulations of DSA \citep[][]{Diesing.Caprioli.22} predict. In future work, we intend to extend our quasilinear treatment to the investigation of relativistic magnetized plasmas that can deepen our understanding of the mechanisms of self-consistent non-thermal particle acceleration. In the context of the solar wind, our assumption of an electrostatic drive is more appropriate for the heating of the ions rather than the electrons. The electrons, being much lighter and more energetic than the ions are more susceptible to electromagnetic (rather than electrostatic) perturbations (such as whistler and Alfven waves), which are not considered in this paper. Moreover, the electrons are typically more susceptible to collisions than the ions, which is why the electron DF tends to Maxwellianize more readily, leading to steeper high $v$ fall-offs \citep[][]{Stverak.etal.09,Maksimovic.21} than the ion DF. 

The introduction of the effect of electromagnetic perturbations and collisions in our model can qualitatively change our conclusions. However, in cases where only collisionless electrostatic plasmas are concerned (such as plasmas in which two-stream instabilities are dominant), our model predicts that the electron DF should exhibit the universal power-laws obtained here.

\section{acknowledgments}
The authors are thankful to the Kavli Institute of Theoretical Physics (KITP), University of California, Santa Barbara, where much of the manuscript was prepared, and to the organizers and attendees of the workshop, ``Interconnections between the Physics of Plasmas and Self-gravitating Systems" at KITP, for insightful discussions. The authors are also thankful to the anonymous referee for insightful comments, and to Toby Adkins, Frank van den Bosch, Pierre-Henri Chavanis, Rebecca Diesing, Robert Ewart, Siddhartha Gupta, Chris Hamilton, Matthew Kunz, Michael Nastac, Alex Schekochihin, Anatoly Spitkovsky, Jonathan Squire, Martin Weinberg, and Vladimir Zhdankin for enlightening discussions and valuable suggestions. This research is supported by the National Science Foundation Award 2209471 and Princeton University.


\bibliography{references_banik}{}
\bibliographystyle{aasjournal}

\appendix

\section{Perturbative response theory for collisionless plasmas: detailed calculations}\label{App:perturbation_theory}

Perturbing the Vlasov-Poisson equations up to linear order, we obtain the evolution equations for the linear order perturbation in the DF, $f_1$ (which we shall henceforth refer to as the linear response), and that in the electric field, $\bE_1$. These are given by the following linearized form of the Vlasov-Poisson equations:

\begin{align}
&\frac{\partial f_1}{\partial t} + \bv\cdot{\bf \nabla}f_1 = -\frac{e}{m}{\bf \nabla}_{\bv} f_0 \cdot \left(\bE^{(\rmP)}+\bE_1\right),\nonumber\\ 
&\nabla\cdot\bE_1 = \frac{e}{\epsilon_0} \int \rmd^d v f_1,
\label{lin_BE_Poisson}
\end{align}
with $d$ the number of dimensions. Similarly, the evolution equations for the second order perturbations, $f_2$ and $\bE_2$, are given by

\begin{align}
&\frac{\partial f_2}{\partial t} + \bv\cdot{\bf \nabla}f_2 + \frac{e}{m}{\bf \nabla}_{\bv} f_0 \cdot \bE_2 = -\frac{e}{m}{\bf \nabla}_{\bv} f_1 \cdot \left(\bE^{(\rmP)}+\bE_1\right),\nonumber\\
&\nabla\cdot\bE_2 = \frac{e}{\epsilon_0} \int \rmd^d v f_2.
\label{2nd_BE_Poisson}
\end{align}

The above equations can be considerably simplified by taking the Fourier transform in $\bx$, i.e., by expanding each quantity as:

\begin{align}
Q_i(\bx,\bv,t) &= \int \rmd^d k\, \exp{\left[i\bk\cdot\bx\right]}\, Q_{i\bk}(\bv,t),
\end{align}
where $i=1,2$ is the order of the perturbation, and the quantity $Q_i$ is equal to $f_i$, $\bE_i$ or $\bE^{(\rmP)}$. The evolution of the first and second order Fourier coefficients is given by the following equations:

\begin{widetext}
\begin{align}
&\frac{\partial f_{1\bk}}{\partial t} + i \bk \cdot \bv f_{1\bk} = -\frac{e}{m}{\bf \nabla}_{\bv} f_0 \cdot \left(\bE^{(\rmP)}_{\bk}+\bE_{1\bk}\right),\nonumber\\
&i\bk\cdot \bE_{1\bk} = \frac{e}{\epsilon_0} \int \rmd^d v f_{1\bk},\nonumber\\
&\frac{\partial f_{2\bk}}{\partial t} + i \bk \cdot \bv f_{2\bk} + \frac{e}{m}{\bf \nabla}_{\bv} f_0 \cdot \bE_{2\bk} = -\frac{e}{m}\int \rmd^d k'\, {\bf \nabla}_{\bv} f_{1\bk'} \cdot \left(\bE^{(\rmP)}_{\bk-\bk'}+\bE_{1\bk-\bk'}\right),\nonumber\\
&i\bk\cdot \bE_{2\bk} = \frac{e}{\epsilon_0} \int \rmd^d v f_{2\bk}.
\end{align}
\end{widetext}
Note that the linear perturbation depends on the equilibrium quantities, while the second order perturbation depends on the linear perturbations. These equations can therefore be solved order by order in perturbation.

\subsection{Linear response theory}

The first step towards solving the perturbed Vlasov-Poisson equations is to solve the linear equations given by equations~(\ref{lin_BE_Poisson}). These are further simplified by taking the Laplace transform in $t$ (not a Fourier transform since we are interested in an initial value problem in the same spirit as \cite{Landau.46}), i.e., by expanding each quantity as:

\begin{align}
Q_{i\bk}(\bv,t) &= \frac{1}{2\pi}\oint \rmd \omega\, \exp{\left[-i\omega t\right]}\, \Tilde{Q}_{i\bk}(\bv,\omega),
\end{align}
where $i=1,2$ is the order of the perturbation, $Q_i$ is equal to $f_i$, $\bE_i$ or $\bE^{(\rmP)}$, and the complex contour integral is performed along the Bromwich contour, i.e., along a loop that consists of the real axis and an infinite radius semicircular arc in the lower half of the complex plane, so that the integral converges.

The linear equations can be solved to yield the following expressions for the Fourier-Laplace coefficients of $f_1$ and $\bE_1$ (with the initial condition that $f_1(t=0)=0$):

\begin{align}
\Tilde{f}_{1\bk}(\bv,\omega) &= -\frac{ie}{m} \, \frac{\left(\Tilde{\bE}^{(\rmP)}_{\bk}(\omega) + \Tilde{\bE}_{1\bk}(\omega)\right)\cdot {\partial f_0}/{\partial \bv}}{\omega - \bk\cdot\bv} ,\nonumber\\
i\bk\cdot \Tilde{\bE}_{1\bk}(\omega) &= \frac{e}{\epsilon_0} \int \rmd^d v \, \Tilde{f}_{1\bk}(\bv,\omega).
\end{align}
Simultaneously solving these equations yields

\begin{align}
\Tilde{\bE}_{\bk}(\omega) &= \Tilde{\bE}^{(\rmP)}_{\bk}(\omega) + \Tilde{\bE}_{1\bk}(\omega) = \frac{\Tilde{\bE}^{(\rmP)}_{\bk}(\omega)}{\varepsilon_{\bk}(\omega)},\nonumber\\
\varepsilon_{\bk}(\omega) &= 1 + \frac{\omega^2_\rmP}{k^2} \int \rmd^d v\, \frac{\bk\cdot{\partial f_0}/{\partial \bv}}{\omega - \bk\cdot\bv}\nonumber\\
&= 1 + \frac{\omega^2_\rmP}{k} \int \rmd^d v\, \frac{{\partial F_0}/{\partial \bv}}{\omega - k v},
\label{lin_resp_eq_app}
\end{align}
where $\omega_\rmP = \sqrt{n_e e^2/m\epsilon_0}$ is the plasma frequency, $n_e$ being the number density of the charged species, and \begin{align}
F_0(v) = \prod_{i=2}^{d} \int \rmd v_i\, f_0(\bv)
\end{align}
is the one-dimensional DF.

The above linear response equation encodes the response of the system to an external drive/perturber. The response-coefficient is the inverse of the dielectric constant, $\varepsilon_{\bk}$, which is a functional of the equilibrium DF, $f_0$. The response therefore depends on the spatio-temporal nature of the perturber. Since the response to a sinusoidal perturber is easy to compute, let us, for the sake of simplicity, rewrite the perturber field as a collection of sinusoids:

\begin{align}
\bE^{(\rmP)}_{\bk}(t) = \int \rmd \omega^{(\rmP)} \,\exp{\left[-i\omega^{(\rmP)}t\right]}\, \bA_{\bk}\left(\omega^{(\rmP)}\right),
\end{align}
whose Laplace transform is given by

\begin{align}
\Tilde{\bE}^{(\rmP)}_{\bk}(\omega) = i\int \rmd \omega^{(\rmP)} \, \frac{\bA_{\bk}\left(\omega^{(\rmP)}\right)}{\omega - \omega^{(\rmP)}}.
\end{align}
Substituting this in the first of equations~(\ref{lin_resp_eq_app}) and taking the inverse Laplace transform yields the following forms for $\bE_{\bk}(t)$ and $f_{1\bk}(t)$:

\begin{widetext}
\begin{align}
\bE_{\bk}(t) &= \int \rmd \omega^{(\rmP)}\, \bA_{\bk}\left(\omega^{(\rmP)}\right) \left[\frac{\exp{\left[-i\omega^{(\rmP)}t\right]}}{\varepsilon_{\bk}\left(\omega^{(\rmP)}\right)} + \sum_n \frac{\exp{\left[-i\omega_{\bk n}t\right]}}{\varepsilon'_{\bk}\left(\omega_{\bk n}\right) \left(\omega_{\bk n} - \omega^{(\rmP)}\right)} \right],\nonumber\\ 
f_{1\bk}(\bv,t) &= -\frac{ie}{m} \frac{\partial f_0}{\partial \bv} \cdot \int \rmd \omega^{(\rmP)} \, \bA_{\bk}\left(\omega^{(\rmP)}\right)\nonumber\\
&\times \left[\frac{1}{\left(\omega^{(\rmP)}-\bk\cdot\bv\right)}\left(\frac{\exp{\left[-i\omega^{(\rmP)}t\right]}}{\varepsilon_{\bk}\left(\omega^{(\rmP)}\right)} - \frac{\exp{\left[-i\bk\cdot \bv\, t\right]}}{\varepsilon_{\bk}\left(\bk\cdot\bv\right)}\right) + \sum_n \frac{\exp{\left[-i\omega_{\bk n}t\right]}}{\varepsilon'_{\bk}\left(\omega_{\bk n}\right) \left(\omega_{\bk n}-\omega^{(\rmP)}\right) \left(\omega_{\bk n}-\bk\cdot\bv\right)}\right],
\label{Ek_f1k_app}
\end{align}
\end{widetext}
where $\omega_{\bk n}$ ($n=0,1,2,...$) are the frequencies of the Landau modes, which are coherent oscillations of the system that follow the Landau dispersion relation, $\varepsilon_{\bk}\left(\omega_{\bk n}\right)=0$ \citep[][]{Landau.46}.

The above linear response consists of three different terms: the free streaming of `dressed' particles that scales as $\sim \exp{\left[-i\bk\cdot\bv t\right]}$, their forced response to the perturber, scaling as $\sim \exp{\left[-i\omega^{(\rmP)}t\right]}$, and the collective excitations or Landau modes, each of which scales as $\sim \exp{\left[-i\omega_{\bk n}t\right]}$. In the stable regime, the Landau modes are all damped (${\rm Im}\, \omega_{\bk n} < 0$), which occurs for $\partial F_0/\partial v < 0$. This implies that, while at times smaller than the damping time of the most weakly damped mode, all three terms contribute to the response, the Landau modes damp away on longer timescales, and the long term response consists of only free streaming and external forcing. In the unstable regime (${\rm Im}\, \omega_{\bk n} > 0$ for at least one mode), which occurs when $\partial F_0/\partial v > 0$ for some $v$, the unstable mode in the third term of the above response dominates on long timescales. 

When the system is in the stable regime, the long term linear response of the system is given by

\begin{widetext}
\begin{align}
\bE_{\bk}(t) &\approx \int \rmd \omega^{(\rmP)}\, \bA_{\bk}\left(\omega^{(\rmP)}\right) \frac{\exp{\left[-i\omega^{(\rmP)}t\right]}}{\varepsilon_{\bk}\left(\omega^{(\rmP)}\right)},\nonumber\\ 
f_{1\bk}(\bv,t) &\approx -\frac{ie}{m} \frac{\partial f_0}{\partial \bv} \cdot \int \rmd \omega^{(\rmP)} \, \frac{\bA_{\bk}\left(\omega^{(\rmP)}\right)}{\left(\omega^{(\rmP)}-\bk\cdot\bv\right)}\left(\frac{\exp{\left[-i\omega^{(\rmP)}t\right]}}{\varepsilon_{\bk}\left(\omega^{(\rmP)}\right)} - \frac{\exp{\left[-i\bk\cdot \bv\, t\right]}}{\varepsilon_{\bk}\left(\bk\cdot\bv\right)}\right).
\label{Ek_f1k_long_time_app}
\end{align}
\end{widetext}
These are the essential ingredients for the computation of the second order/quasilinear response, which we discuss next.

\subsection{Quasilinear response theory}\label{App:QLT}

We want to investigate the evolution of the mean DF averaged over the volume $V$ of the bulk plasma,
\begin{align}
f_0\left(\bv,t\right) &= \frac{\int \rmd^d x \, f\left(\bx,\bv,t\right)}{V} \nonumber\\
&\approx \frac{1}{V} \int \rmd^d x \int \rmd^d k\, \exp{\left[i\bk\cdot\bx\right]}\, f_{2\bk}\left(\bv,t\right) \nonumber\\
&= \frac{1}{V} \int \rmd^d k\, f_{2\bk}\left(\bv,t\right) \int \rmd^d x\, \exp{\left[i\bk\cdot\bx\right]} \nonumber\\
&= \frac{{\left(2\pi\right)}^d}{V} \int \rmd^d k\, f_{2\bk}\left(\bv,t\right)\, \delta^d\left(\bk\right) \nonumber\\
&= \frac{{\left(2\pi\right)}^d}{V} f_{2\bk=0}\left(\bv,t\right),
\end{align}
where the approximation symbol denotes the quasilinear/second order approximation (note that the spatially averaged linear order perturbation is zero since $f_{1\bk}\propto \bk\cdot\partial f_0/\partial \bv \to 0$ as $\bk\to 0$). In going from the third to the fourth line, we have used the identity that $\int \rmd^d x\,e^{i\bk\cdot\bx} = {\left(2\pi\right)}^d \delta^d\left(\bk\right)$.

The evolution of the mean DF, $f_0 = {\left(2\pi\right)}^d f_{2\bk = 0}/V$, is given by the following quasilinear equation, which is obtained by taking the $\bk\to 0$ limit of the evolution equation for $f_{2\bk}$ \citep[][]{Diamond_Itoh_Itoh_2010}:

\begin{align}
\frac{\partial f_0}{\partial t} &= -\frac{{\left(2\pi\right)}^d e}{m V} \int \rmd^d k \, \left<\bE_{\bk}^{\ast} \cdot \nabla_{\bv} f_{1\bk}\right>.
\label{quasilin_resp_eq_app}
\end{align}
Here we have used the reality condition, $\bE_{-\bk}^{(\rmP)} = \bE_{\bk}^{(\rmP)\ast}$. 

Now we need to make assumptions about $A_{\bk}\left(\omega^{(\rmP)}\right)$, i.e., about the external electric field, $E_{\bk j}^{(\rmP)}(t)$, where the subscript $j$ denotes the $j^{\rm th}$ component of $\bE_{\bk}^{(\rmP)}(t)$. For simplicity, we assume that the $E_{\bk j}^{(\rmP)}(t)$ is a red noise of the following form:

\begin{align}
\left<E_{\bk j}^{(\rmP)\ast}(t) E_{\bk l}^{\rmP}(t')\right> &= \calE_{jl}\left(\bk\right)\,\calC_t\left(t-t'\right).
\label{white_noise_t_app}
\end{align}
This implies that

\begin{widetext}
\begin{align}
\left<A_{\bk j}^\ast\left(\omega^{(\rmP)}\right) A_{\bk l}\left(\omega^{'(\rmP)}\right)\right> &= \frac{1}{{\left(2\pi\right)}^2} \int \rmd t \int \rmd t' \exp{\left[i\left(\omega^{(\rmP)}t - \omega^{'(\rmP)}t'\right)\right]} \left<E_{\bk j}^{(\rmP)\ast}(t) E_{\bk l}^{\rmP}(t')\right> \nonumber\\
&= \calE_{jl}\left(\bk\right)\, \calC_{\omega}\left(\omega^{(\rmP)}\right) \delta\left(\omega^{(\rmP)}-\omega^{'(\rmP)}\right),
\label{white_noise_omega_app}
\end{align}
\end{widetext}
where 

\begin{align}
\calC_{\omega}\left(\omega^{(\rmP)}\right) &= \frac{1}{2\pi} \int \rmd t \exp{\left[-i\omega^{(\rmP)} t\right]} \, \calC_t\left(t\right).
\end{align}

Substituting the expressions for $\bE_{\bk}(t)$ and $f_{1\bk}(\bv,t)$ from equations~(\ref{Ek_f1k_app}) in the above quasilinear equation~(\ref{quasilin_resp_eq_app}) and using the red noise condition for the perturbing electric field given in equation~(\ref{white_noise_t_app}), we obtain the following equation for the quasilinear relaxation of $f_0$:

\begin{align}
\frac{\partial f_0}{\partial t} &= \frac{\partial}{\partial v_j}\left(D_{jl}(\bv,t)\frac{\partial f_0}{\partial v_l}\right),
\label{quasilin_resp_FP_eq_app}
\end{align}
where $D_{jl}$ is given by

\begin{widetext}
\begin{align}
D_{jl}(\bv,t) &= \frac{i {\left(2\pi\right)}^d e^2}{m^2 V} \int \rmd^d k\, \calE_{jl}(\bk) \left[\int \rmd\omega^{(\rmP)} \,\calC_{\omega}\left(\omega^{(\rmP}\right) \, \left[\frac{1}{\left(\omega^{(\rmP)}-\bk\cdot\bv\right)\varepsilon^\ast_{\bk}\left(\omega^{(\rmP)}\right)} \left[\frac{1}{\varepsilon_{\bk}\left(\omega^{(\rmP)}\right)} - \frac{\exp{\left[i\left(\omega^{(\rmP)}-\bk\cdot v\right)t\right]}}{\varepsilon_{\bk}(\bk\cdot\bv)}\right]\nonumber\right.\right.\\
&\left.\left.+\sum_{n,p}\frac{\exp{\left[\left(\gamma_{\bk n}+\gamma_{\bk p}\right)t + i\left(\eta_{\bk n}-\eta_{\bk p}\right)t\right]}}{\varepsilon^{'\ast}_{\bk}\left(\omega_{\bk n}\right) \varepsilon'_{\bk}\left(\omega_{\bk p}\right)\left(\omega^{(\rmP)}-\omega_{\bk p}^{\ast}\right)\left(\omega^{(\rmP)}-\omega_{\bk n}\right)\left(\omega_{\bk n}-\bk\cdot\bv\right)}\right]\right],
\label{diff_coeff_full_app}
\end{align}
\end{widetext}
where $\omega_{\bk n} = \eta_{\bk n} + i\gamma_{\bk n}$. Equation~(\ref{quasilin_resp_FP_eq_app}) is nothing but a Fokker-Planck equation with the diffusion tensor given above. The diffusion tensor consists of two terms: the first term stands for the direct interaction between the perturber and the dressed particles, while the second term represents wave-wave interactions. In the stable regime, i.e., when all the Landau modes are damped, both terms contribute to the diffusion coefficient at times smaller than the damping timescale of the least damped Landau mode. At longer times, after the Landau modes have damped away, only the external forcing contributes to diffusion. In the unstable regime, which corresponds to $\partial F_0/\partial v >0$, the unstable modes of the wave-wave term dominate at long time.

In the stable regime, at times smaller than the Landau damping time, for which we can take the $\gamma_{\bk n}\to 0$ and $t\to \infty$ limit, $D_{jl}$ becomes

\begin{align}
D_{jl}(\bv) &= \frac{2^d \pi^{d+1} e^2}{m^2 V} \int \rmd^d k\, \calE_{jl}(\bk)\, \calC_{\omega}\left(\bk\cdot\bv\right) \nonumber\\
&\times \left(\frac{1}{{\left|\varepsilon_{\bk}\left(\bk\cdot\bv\right)\right|}^2} + \frac{1}{{\left(\bk\cdot\bv - \eta_{\bk}\right)}^2 {\left|\varepsilon'_{\bk}\left(\eta_{\bk}\right)\right|}^2}\right),
\end{align}
where we have used the identity that $\lim_{t\to\infty}\exp{\left[ix t\right]}/x = 1/x + i\pi \delta(x)$. In the long time limit, the Landau modes damp away, and only the first term survives, which yields

\begin{align}
D_{jl}(\bv) &\approx \frac{2^d \pi^{d+1} e^2}{m^2 V} \int \rmd^d k\, \frac{\calE_{jl}(\bk)\, \calC_{\omega}\left(\bk\cdot\bv\right)}{{\left|\varepsilon_{\bk}\left(\bk\cdot\bv\right)\right|}^2}.
\end{align}
\\
\section{Computation of the quasilinear diffusion coefficient}\label{App:diff_coeff}

\subsection{White noise}\label{App:diff_coeff_wn}

In the case of the white noise, where $\calC_t\left(t-t'\right)$ is equal to $\delta\left(t-t'\right)$, we have $\calC_{\omega}\left(\bk\cdot\bv\right)=1$, and the diffusion coefficient simplifies to the following (assuming isotropy and $d=3$):

\begin{align}
&D(v) = \frac{32\pi^5 e^2}{m^2 V} \int_0^{\infty} \rmd k\, k^2 \calE(k) \int_{0}^{1}\rmd\left(\cos{\theta}\right) \frac{1}{{\left|\varepsilon_{k}\left(kv\cos\theta\right)\right|}^2}.
\label{diff_coeff_app}
\end{align}
Its functional form solely depends on that of the dielectric constant. Therefore, it is instructive to take a look at the velocity dependence of $\varepsilon_k\left(k v \cos{\theta}\right)$. For $v \lesssim \omega_\rmP/k$, we can Taylor expand the principal value in $v'/v\cos{\theta}$ or $v\cos{\theta}/v'$ (depending on which is smaller than unity) and truncate up to second order to obtain the following approximate expression for $\varepsilon_k$:

\begin{widetext}
\begin{align}
\varepsilon_k(kv\cos\theta) &\approx 1-\frac{\omega^2_\rmP}{k^2 v^2 \cos^2{\theta}}\left[1 + \frac{6}{v^2\cos^2\theta}\int_{0}^{v\cos{\theta}} \rmd v' v'^2 F_0(v) \right] \nonumber\\
&-\frac{2 \omega^2_\rmP}{k^2} \left[\int_{v\cos{\theta}}^\infty \rmd v'\, \frac{1}{v'}\frac{\partial F_0}{\partial v'} + v\cos{\theta} \int_{v\cos{\theta}}^\infty \rmd v'\, \frac{1}{v'^2}\frac{\partial F_0}{\partial v'} \right] \nonumber\\
&-i\pi \frac{\omega^2_\rmP}{k^2} \left.\frac{\partial F_0}{\partial v}\right|_{v=\omega_\rmP/k\cos\theta}.
\end{align}
\end{widetext}
Note that $\varepsilon_k$ is approximately equal to $1 - \omega^2_\rmP/\left(k^2 v^2 \cos^2{\theta}\right) - i\pi\left(\omega^2_\rmP/k^2\right) \left.\partial F_0/\partial v\right|_{v=\omega_\rmP/k\cos{\theta}}$ for $\sigma\lesssim v \lesssim \omega_\rmP/k$, and roughly equal to $1 + c_{\rmF}\,\omega^2_\rmP/k^2\sigma^2$ for $v \lesssim \sigma$, with $c_\rmF$ an $\calO(1)$ constant that depends on the high $v$ asymptotic behavior of $F_0$. The imaginary part of $\varepsilon_{k}$ is almost always small for $k\lambda_\rmD \ll 1$. Therefore, $\varepsilon_k$ scales as $\sim 1/v^2$ for $\sigma \lesssim v \lesssim \omega_\rmP/k$, and approaches a constant in the $v\ll \sigma$ limit. Moreover, $\varepsilon_k$ tends to $1$ in the limit of $v \gg \omega_\rmP/k$.  

Using the above behavior of $\varepsilon_k$, we can approximately evaluate the diffusion coefficient for $\sigma \lesssim v \lesssim \omega_\rmP/k$ as follows:

\begin{align}
D(v) &\approx \frac{32\pi^5 e^2 v^4}{m^2 V} \int_0^{\infty} \rmd k\, k^6 \calE(k) \nonumber\\
&\times \int_{0}^{1}\rmd\left(\cos{\theta}\right) \frac{\cos^4{\theta}}{{\left(k^2 v^2 \cos^2{\theta} - \omega^2_\rmP\right)}^2},
\end{align}
which can be integrated to yield

\begin{widetext}
\begin{align}
&D(v) \approx \frac{32\pi^5 e^2}{m^2 V} \int_0^{\infty} \rmd k\, k^2 \calE(k) \left[1 - \frac{3}{4}\frac{\omega_\rmP}{k v}\ln{\left(\left|\frac{\omega_\rmP + kv}{\omega_\rmP - kv}\right|\right)} + \frac{1}{2} \frac{\omega^2_\rmP}{\omega^2_\rmP-k^2 v^2} \right].
\label{D_approx_wn_app}
\end{align}
\end{widetext}
Note that this diverges at $v=\omega_\rmP/k$, since we have neglected the imaginary part of $\varepsilon_{k}$ in the denominator of the integrand. Including this would yield a large but finite answer at $v=\omega_\rmP/k$ (see Fig.~\ref{fig:diff_coeff}). By expanding the above in $kv/\omega_\rmP$, we can see that $D(v)$ scales as $v^4$ for $\sigma \lesssim v \ll \omega_\rmP/k $. The asymptotic scalings of $D(v)$ are summarized in equation~(\ref{diff_coeff_asymptotic_wn}).

\subsection{Red noise}\label{App:diff_coeff_rn}

Let the correlation function, $\calC_t$, be of the form

\begin{align}
\calC_t\left(t-t'\right) = \frac{1}{2 \tau_\rmc} \exp{\left[-\left|t-t'\right|/\tau_\rmc\right]},
\end{align}
for which

\begin{align}
\calC_{\omega}\left(\bk\cdot\bv\right) = \frac{1}{1 + {\left(\bk\cdot\bv \, \tau_\rmc\right)}^2}.
\end{align}
This tends to $1$ as $\tau_\rmc \to 0$, since $\calC_t\left(t-t'\right) \to \delta\left(t-t'\right)$ and the red noise becomes white in this limit. 

When the noise is of the above form, the quasilinear diffusion coefficient becomes

\begin{align}
&D(v) = \frac{32\pi^5 e^2}{m^2 V} \int_0^{\infty} \rmd k\, k^2 \calE(k) \nonumber\\
&\times \int_{0}^{1}\rmd\left(\cos{\theta}\right) \frac{1}{1 + {\left(k v \tau_\rmc \cos{\theta}\right)}^2} \frac{1}{{\left|\varepsilon_{k}\left(kv\cos\theta\right)\right|}^2},
\end{align}
Using the approximate form of $\varepsilon_k$ for $\sigma \lesssim v \lesssim \omega_\rmP/k$ as discussed in Appendix~\ref{App:diff_coeff_wn}, this can be written as

\begin{align}
&D(v) \approx \frac{32\pi^5 e^2 v^4}{m^2 V} \int_0^{\infty} \rmd k\, k^6 \calE(k) \nonumber\\
&\times \int_{0}^{1}\rmd\left(\cos{\theta}\right) \frac{1}{1 + {\left(k v \tau_\rmc \cos{\theta}\right)}^2} \frac{\cos^4{\theta}}{{\left(k^2 v^2 \cos^2{\theta} - \omega^2_\rmP\right)}^2},
\end{align}
in the range $\sigma \lesssim v \lesssim \omega_\rmP/k$. The $\cos{\theta}$ integral can be performed to yield the following form for $D(v)$:

\begin{widetext}
\begin{align}
D(v) &\approx \frac{32 \pi^5 e^2}{m^2 V \tau_\rmc v} \int_0^{\infty} \rmd k\, k\, \calE(k) \left[\frac{\tan^{-1}\left(k v \tau_\rmc\right)}{{\left(1+{\omega_\rmP^2 \tau_\rmc^2}\right)}^2} - \frac{\omega_\rmP \tau_\rmc}{4}\frac{3+{\omega_\rmP^2\tau_\rmc^2}}{{\left(1+{\omega_\rmP^2 \tau_\rmc^2}\right)}^2}\ln{\left(\left|\frac{\omega_\rmP + k v}{\omega_\rmP - k v}\right|\right)} + \frac{1}{2}\frac{k v \tau_\rmc}{1+{\omega_\rmP^2 \tau_\rmc^2}}\frac{\omega^2_\rmP}{\omega^2_\rmP-{k^2 v^2}} \right],
\label{D_approx_rn1_app}
\end{align}
\end{widetext}
which, for $\omega_\rmP \tau_\rmc \gtrsim 1$ and $v<\omega_\rmP/k$, reduces to

\begin{align}
D(v) &\approx \frac{32 \pi^5 e^2}{m^2 V \omega^4_\rmP \tau^5_\rmc \, v} \int_0^{\infty} \rmd k\, k\, \calE(k) \nonumber\\
&\times \left[\tan^{-1}\left(k v \tau_\rmc\right) - k v\tau_\rmc + \frac{{\left(k v \tau_\rmc\right)}^3}{3} \right].
\label{D_approx_rn2_app}
\end{align}
The asymptotic scalings of $D(v)$ are summarized in equation~(\ref{diff_coeff_asymptotic_rn}).

\section{Solution of the quasilinear diffusion equation}\label{App:self_similar_sol_deriv}

\subsection{Numerical solution for $d>1$}\label{App:code}

Here we chalk out the details of the finite difference flux-conserving code used to numerically integrate the quasilinear diffusion equation~(\ref{QL_diffusion_eqn_nd}) for $d>1$. We discretize the velocity range, $\left(u_{\rm min},u_{\rm max}\right)$, into $N_u$ grid cells so that $\Delta u = \left(u_{\rm max}-u_{\rm min}\right)/\left(N_u-1\right)$ is the grid size. We use $\Delta \tau$ to denote the temporal step size, $i$ to index the timestep, and $j$ to index the velocity cell. The diffusive fluxes of particles leaving and entering the $j^{\rm th}$ velocity cell are $F_{j+1/2}^i$ and $F_{j-1/2}^i$, respectively given by

\begin{align}
F_{j+1/2}^i = {\left(u^{d-1} \Tilde{D}\right)}_{j+1/2}^i \frac{f_{0,j+1}^i-f_{0,j}^i}{\Delta u},\nonumber\\
F_{j-1/2}^i = {\left(u^{d-1} \Tilde{D}\right)}_{j-1/2}^i \frac{f_{0,j}^i-f_{0,j-1}^i}{\Delta u},
\end{align}
where ${\left(u^{d-1} \Tilde{D}\right)}_{j+1/2}^i$ and ${\left(u^{d-1} \Tilde{D}\right)}_{j-1/2}^i$ are given by the following harmonic means of the quantities evaluated at adjacent cells:

\begin{align}
{\left(u^{d-1} \Tilde{D}\right)}_{j+1/2}^i = \frac{{\left(u^{d-1} \Tilde{D}\right)}_{j}^i\, {\left(u^{d-1} \Tilde{D}\right)}_{j+1}^i}{{\left(u^{d-1} \Tilde{D}\right)}_{j}^i + {\left(u^{d-1} \Tilde{D}\right)}_{j+1}^i},\nonumber\\
{\left(u^{d-1} \Tilde{D}\right)}_{j-1/2}^i = \frac{{\left(u^{d-1} \Tilde{D}\right)}_{j}^i\, {\left(u^{d-1} \Tilde{D}\right)}_{j-1}^i}{{\left(u^{d-1} \Tilde{D}\right)}_{j}^i + {\left(u^{d-1} \Tilde{D}\right)}_{j-1}^i}.
\end{align}
The quasilinear diffusion equation~(\ref{QL_diffusion_eqn_nd}) can then be rewritten in the following discretized format:

\begin{align}
\frac{f_{0,j}^{i+1}-f_{0,j}^{i}}{\Delta\tau} = \frac{1}{u_j^{d-1}} \frac{F_{j+1/2}^i - F_{j-1/2}^i}{\Delta u}.
\end{align}

We implement the boundary condition that the incoming flux $F_{-1/2}^i$ into the velocity range and the outgoing flux $F_{N_u - 1/2}^i$ from the velocity range are both zero, which boils down to the Neumann boundary condition that $\partial f_0/\partial u=0$ at both ends.

To ensure the numerical stability of the code, we follow the Courant condition that

\begin{align}
\Delta t < \frac{{\left(\Delta u\right)}^2}{2 \Tilde{D}_{\rm max}},
\end{align}
where $\Tilde{D}_{\rm max}$ is the maximum value of the diffusion coefficient $\Tilde{D}$ over the entire velocity range.

\subsection{Self-similar solution for $d>1$}\label{App:self_similar_sol}

The quasilinear diffusion coefficient has a non-trivial dependence on $v$, but, as shown in the previous section, can be written as a combination of different power-laws, i.e., $D(v)\sim v^\alpha$ with a different $\alpha$ in a different range, especially for forcing on scales larger than the Debye length (see equations~[\ref{diff_coeff_asymptotic_wn}] and [\ref{diff_coeff_asymptotic_rn}]). Considering a white noise drive, we have $\alpha=0$ for $v\lesssim \sigma$ and $v\gtrsim \left(1/k\lambda_\rmD\right)\left(\omega_\rmP/k\right)$, $\alpha=-1$ for $\omega_\rmP/k\lesssim v \lesssim \left(1/k\lambda_\rmD\right)\left(\omega_\rmP/k\right)$, and $\alpha = 4$ for $\sigma \lesssim v \lesssim \omega_\rmP/k$. For a red noise drive of the form given in equation~(\ref{red_noise_model}) with $\omega_\rmP \tau_\rmc > 1 \gtrsim k\sigma \tau_\rmc$, $\alpha=4$ for $\sigma \lesssim v \lesssim 1/k\tau_\rmc$ but $\approx 2$ for $1/k\tau_\rmc \lesssim v \lesssim \omega_\rmP/k$ and $-1$ beyond. All in all, $\alpha = 4$ over a large range of velocities for a sufficiently large-scale, white noise-like forcing. 

The power-law form of the diffusion coefficient implies the existence of self-similar solutions to the Fokker-Planck equation given in equation~(\ref{QL_diffusion_eqn_nd}). Let us therefore try the following ansatz: $f_0(u,\tau) = \tau^a \Psi(\xi)$ with $\xi = u/\tau^b$. We have to solve for $a$ and $b$ in terms of $\alpha$ and $d$. This requires us to solve two equations. Besides the diffusion equation, we solve an equation for particle number conservation, i.e., we ensure that the following is approximately true:

\begin{align}
\int \rmd u\, u^{d-1} f_0(u) = {\rm constant}
\label{particle_no_conserv_eqn_app}
\end{align}
in each velocity range corresponding to a single power-law. Note that this approximation strictly holds only in a quasi-steady state, where there is a constant flux of particles into and out of each velocity interval. Substituting $f_0(u,\tau) = \tau^a \Psi(\xi)$ with $\xi = u/\tau^b$ in equations~(\ref{QL_diffusion_eqn_nd}) and (\ref{particle_no_conserv_eqn_app}), and solving the two resultant equations for $a$ and $b$ in terms of $\alpha$ and $d$, we obtain

\begin{align}
a = \frac{d}{\alpha - 2},\;
b = -\frac{1}{\alpha - 2}.
\end{align}
We also obtain the following second order ODE in $\xi$:

\begin{align}
\frac{\rmd}{\rmd \xi}\left(\xi^{\alpha+d-1}\frac{\rmd \Psi}{\rmd \xi}\right) &= \frac{1}{\alpha - 2}\frac{\rmd}{\rmd \xi}\left(\xi^d \Psi\right),
\end{align}
which can be integrated once to obtain the following first order ODE

\begin{align}
\frac{\rmd \Psi}{\rmd \xi} - \frac{\xi^{1-\alpha}}{\alpha - 2}\Psi = c_1 \xi^{-\left(\alpha+d-1\right)},
\end{align}
with $c_1$ an integration constant. We can integrate it once more using the method of integrating factor to obtain the following solution for $\Psi$:

\begin{align}
\Psi\left(\xi\right) &= c_1 \exp{\left[-\xi^{2-\alpha}/{\left(2-\alpha\right)}^2\right]} \nonumber\\
&\times \int \rmd \xi' \exp{\left[\xi'^{\,2-\alpha}/{\left(2-\alpha\right)}^2\right]}\xi'^{-\left(\alpha+d-1\right)} \nonumber\\
&+ c_2 \exp{\left[-\xi^{2-\alpha}/{\left(2-\alpha\right)}^2\right]}.
\end{align}
Now we employ the boundary conditions that $\Psi\to 0$ as $\xi \to \infty$ (i.e., $v\to \infty$) and $\int_{\xi} \rmd \xi' \, \xi'^{d-1} \Psi(\xi')$ is finite as $\xi \to 0$ (i.e., $v\to 0$). This fixes $c_1$ and $c_2$, allowing the following class of solutions:

\begin{widetext}
\begin{align}
\Psi\left(\xi\right) &= \exp{\left[-\xi^{2-\alpha}/{\left(2-\alpha\right)}^2\right]} \; \times 
&\begin{cases}
1,\qquad \qquad \qquad \qquad \qquad \qquad \qquad \qquad \qquad \;\;\, \xi < \tau^{\frac{1}{\alpha-2}},  \\
\int_{\xi}^\infty \rmd \xi' \exp{\left[\xi'^{\,2-\alpha}/{\left(2-\alpha\right)}^2\right]}\xi'^{-\left(\alpha+d-1\right)}, \qquad \qquad \;\, \tau^{\frac{1}{\alpha-2}} < \xi < \frac{1}{k\lambda_\rmD}\tau^{\frac{1}{\alpha-2}}, \\
1, \qquad \qquad \qquad \qquad \qquad \qquad \qquad \qquad \qquad \;\; \frac{1}{k\lambda_\rmD} \tau^{\frac{1}{\alpha-2}} < \xi < \frac{1}{{\left(k\lambda_\rmD\right)}^2} \tau^{\frac{1}{\alpha-2}}, \\
1, \qquad \qquad \qquad \qquad \qquad \qquad \qquad \qquad \qquad \;\;\, \xi > \frac{1}{{\left(k\lambda_\rmD\right)}^2} \tau^{\frac{1}{\alpha-2}}.
\end{cases}
\label{Psi_sol}
\end{align}
\end{widetext}
Here, $\alpha=0$ in the first and fourth intervals, $4$ in the second and $-1$ in the third interval. It is instructive to look at the asymptotic behaviour of $\Psi(\xi)$. Substituting the value of $\alpha$ appropriate for each interval in the above equation, and taking the asymptotic limits, $\xi \to 0$ and/or $\xi \to \infty$ in each case, we obtain the following scalings for $\Psi(\xi)$ in the case of white noise forcing:

\begin{widetext}
\begin{align}
\Psi(\xi) &\approx
\begin{cases}
\exp{\left[-\xi^2/4\right]}, \qquad \qquad \quad \qquad \qquad \qquad \xi \lesssim \tau^{-\frac{1}{2}}, \\
2\, \xi^{-d}, \qquad \qquad \qquad \qquad \qquad \qquad \quad \;\;\, \tau^{-\frac{1}{2}} \lesssim \xi \lesssim \max{\left[1,\tau^{\frac{1}{2}}\right]}, \\
\dfrac{\xi^{-\left(d+2\right)}}{d+2}, \qquad \qquad \qquad \qquad \qquad \qquad \;\;\; \max{\left[1,\tau^{\frac{1}{2}}\right]} \lesssim \xi \lesssim {\dfrac{1}{k\lambda_\rmD}}\tau^{1/2}, \\
\exp{\left[-\xi^3/9\right]}, \qquad \qquad \qquad \qquad \qquad \;\;\; \dfrac{1}{k\lambda_\rmD}\tau^{1/2} \lesssim \xi \lesssim \dfrac{1}{{\left(k\lambda_\rmD\right)}^2} \tau^{1/2}, \\
\exp{\left[-\xi^2/4\right]}, \qquad \qquad \qquad \qquad \qquad \quad \xi \gtrsim \dfrac{1}{{\left(k\lambda_\rmD\right)}^2} \tau^{1/2}.
\end{cases}
\end{align}
\end{widetext}
Noting that $f_0 = t^a \Psi(\xi)$ and $a=d/\left(\alpha-2\right)$, we obtain the following dependencies of $f_0$ on $v$ and $t$:

\begin{widetext}
\begin{align}
f_0(v,t) &\sim
\begin{cases}
\sigma^{-d}_1(t) \exp{\left[-v^2/2\sigma^2_1(t)\right]} \qquad \qquad \qquad\, v \lesssim \sigma_1, \\
\dfrac{v^{-\left(d+2\right)}}{d+2} t^{-1}, \qquad \qquad \qquad \qquad \qquad \quad \, \sigma_1 \lesssim v \lesssim {\dfrac{\omega_\rmP}{k}},\\
\sigma^{-2d/3}_2(t) \exp{\left[-\frac{\sqrt{8}}{9}{\left(v^2/2\sigma^2_2(t)\right)}^{3/2}\right]}, \quad \;\, \dfrac{\omega_\rmP}{k} \lesssim v \lesssim \dfrac{1}{k\lambda_\rmD} \dfrac{\omega_\rmP}{k},\\
\sigma^{-d}_2(t) \exp{\left[-v^2/2\sigma^2_2(t)\right]} \qquad \qquad \qquad \, v \gtrsim \dfrac{1}{k\lambda_\rmD} \dfrac{\omega_\rmP}{k},
\end{cases}
\end{align}
\end{widetext}
with

\begin{align}
&\sigma^2_1(t) = \sigma^2 + 2D_0{\left(k\lambda_\rmD\right)}^4 t,\nonumber\\
&\sigma^2_2(t) \approx {\left(\omega_\rmP/k\right)}^2 + 2 D_0 t,
\end{align}
where $D_0 = 32\pi^5 e^2 k^2 \calE_0 /m^2 V$. Here we have assumed that $\calE(k) = \calE_0\,\delta\left(k-k_0\right)$. For a red noise drive of the form given in equation~(\ref{red_noise_model}) with $\omega_\rmP \tau_\rmc > 1 \gtrsim k\sigma_1 \tau_\rmc$, $f_0$ scales as $v^{-\left(d+2\right)} t^{-1}$ for $\sigma_1 \lesssim v \lesssim 1/k\tau_\rmc$ but roughly as $v^{-d}\,t^{-1}$ for $1/k\tau_\rmc \lesssim v \lesssim \omega_\rmP/k$, since $\alpha$ is approximately $2$ in this interval \footnote{There is a subtle catch here. This is only true in the limiting sense, i.e., $G(v)$ scales as $\sim v^{-{\left(d+\Delta \alpha\right)}}$ for $1/k\tau_\rmc \lesssim v \lesssim \omega_\rmP/k$, where $\alpha=2+\Delta\alpha$ with $\Delta \alpha$ small but positive, so that the particle number does not diverge as $v\to \infty$. If, on the other hand, $\alpha$ is exactly equal to $2$, i.e., $D=D' v^2$, then $G(v)$ scales differently, e.g., as $\sim t^{-1/2} \exp{\left[-9 D'/4 t\right]} \, v^{-3/2} \exp{\left[-{\ln{\left(v/v_0\right)}}^2/4 D' t\right]}$ in 3D, as shown by \citep[][]{Jokipii.Lee.10}.}.

\subsection{Solution in 1D}

In 1D, the quasilinear equation can be written as

\begin{align}
\frac{\partial f_0}{\partial t} = \frac{\partial}{\partial v}\left(D(v)\frac{\partial f_0}{\partial v}\right),
\end{align}
with $D(v)$ given by

\begin{align}
D(v) = \frac{2\pi^2 e^2}{m^2 V} \int \rmd k\, \frac{\calE(k) \, \calC_{\omega}(kv)}{{\left|\varepsilon\left(kv\right)\right|}^2},
\end{align}
and $\varepsilon(kv)$ given by

\begin{align}
\varepsilon_{k}\left(kv\right) &= 1 + \frac{\omega^2_\rmP}{k^2} \int \rmd v'\, \frac{{\partial f_0}/{\partial v'}}{v - v'}.
\end{align}

Let us study what happens for $k\ll k_\rmD$ and $v\approx \omega_\rmP/k$. At these velocities, $\partial f_0/\partial v$ is quite small (especially so for a Maxwellian $f_0$), and we are justified in taking the following limit (assuming white noise, i.e., $\calC_{\omega}\left(kv\right)=1$):

\begin{widetext}
\begin{align}
\lim_{\partial f_0/\partial v \to 0} D(v) \frac{\partial f_0}{\partial v} &= \frac{2\pi^2 e^2}{m^2 V} \int \rmd k\, \calE(k) \lim_{\partial f_0/\partial v \to 0} \frac{\partial f_0/\partial v}{{\left(1-\omega^2_\rmP/k^2 v^2\right)}^2 + {\left(\pi \omega^2_\rmP/ k^2\right)}^2 {\left(\partial f_0/\partial v\right)}^2 } \nonumber\\
&= -\frac{\pi e^2}{m^2 V} \frac{1}{\omega_\rmP} \int \rmd k\, \calE(k)\, k^2\, \delta\left(kv - \omega_\rmP\right) \nonumber\\
&= -\frac{\pi e^2}{m^2 V} \frac{\omega_\rmP}{v^3} \, \calE\left(\frac{\omega_\rmP}{v}\right)
\end{align}
\end{widetext}
Differentiating the above with respect to $v$, we obtain

\begin{align}
&\lim_{\partial f_0/\partial v \to 0} \frac{\partial}{\partial v} \left(D(v) \frac{\partial f_0}{\partial v}\right) \nonumber\\
&= \frac{\pi^2 e^2}{m^2 V} \frac{\omega_\rmP}{v^4} \left[3 \, \calE\left(\frac{\omega_\rmP}{v}\right) + \frac{\omega_\rmP}{v} \calE'\left(\frac{\omega_\rmP}{v}\right)\right].
\end{align}
If $\calE(k) = \calE_0 \, \delta\left(k-k_0\right)$, then the above reduces to

\begin{align}
&\lim_{\partial f_0/\partial v \to 0} \frac{\partial}{\partial v} \left(D(v) \frac{\partial f_0}{\partial v}\right) \nonumber\\
&= \frac{\pi^2 e^2 \calE_0}{m^2 V} \frac{1}{v^2} \left[\delta\left(v - \frac{\omega_\rmP}{k_0}\right) - v\, \delta'\left(v - \frac{\omega_\rmP}{k_0}\right)\right].
\end{align}
This serves as an inhomogeneous source term in the quasilinear diffusion equation, which therefore becomes 

\begin{align}
\frac{\partial f_0}{\partial t} &= \left.\frac{\partial}{\partial v}\left(D(v)\frac{\partial f_0}{\partial v}\right)\right|_{v \neq \omega_\rmP/k} \nonumber\\
&+ \frac{\pi^2 e^2}{m^2 V} \frac{\omega_\rmP}{v^4} \left[3 \, \calE\left(\frac{\omega_\rmP}{v}\right) + \frac{\omega_\rmP}{v} \calE'\left(\frac{\omega_\rmP}{v}\right)\right].
\end{align}
The solution consists of a homogeneous part that behaves like a Maxwellian core for $v<\sigma$ and scales as $v^{-3}$ for $\sigma<v<\omega_\rmP/k$, and an inhomogeneous part. For $\calE(k) = \calE_0 \, \delta\left(k-k_0\right)$, the inhomogeneous part consists of a Dirac delta spike, which renders the DF linearly unstable at $v=\omega_\rmP/k$. Now, another term proportional to $\exp{\left[2\gamma t\right]}$, where $\gamma$ is the growth rate of the unstable Landau mode, appears in the quasilinear diffusion equation (this arises from the second term in equation~[\ref{diff_coeff_full_app}]). Ultimately, this term saturates the instability and forms a plateau around $v=\omega_\rmP/k$. This plateau is more pronounced in 1D than in higher dimensions. This is because resonant heating is more pronounced in 1D; in higher dimensions, for isotropic forcing as we assume in this paper, the Dirac delta spike in the diffusion coefficient around $v=\omega_\rmP/k$ is broadened due to marginalization over the solid angle.


\label{lastpage}



\end{document}